\documentclass[%
 reprint,
 amsmath,amssymb,
 aps,usenames,dvipsnames
]{revtex4-2}

\usepackage{natbib}
\usepackage{graphicx}
\usepackage{dcolumn}
\usepackage{bm}
\usepackage{ulem}
\usepackage{hyperref}
\usepackage{cleveref}

\usepackage[table,dvipsnames]{xcolor}

\definecolor{brightgreen}{rgb}{0.4, 1.0, 0.0}
\definecolor{teagreen}{rgb}{0.82, 0.94, 0.75}
\definecolor{springgreen}{rgb}{0.0, 1.0, 0.5}
\definecolor{persiangreen}{rgb}{0.0, 0.65, 0.58}
\definecolor{cherryblossompink}{rgb}{1.0, 0.72, 0.77}
\definecolor{mediumspringgreen}{rgb}{0.0, 0.98, 0.6}
\definecolor{bittersweet}{rgb}{1.0, 0.44, 0.37}
\definecolor{cinnabar}{rgb}{0.89, 0.26, 0.2}
\definecolor{applegreen}{rgb}{0.55, 0.71, 0.0}
\definecolor{caribbeangreen}{rgb}{0.0, 0.8, 0.6}
\definecolor{turquoisegreen}{rgb}{0.63, 0.84, 0.71}
\definecolor{ufogreen}{rgb}{0.24, 0.82, 0.44}

\begin{document}

\preprint{APS/123-QED}

\title{Atomistic transport modeling, design principles and empirical rules for Low Noise III-V Digital Alloy Avalanche Photodiodes}

\author{Sheikh Z. Ahmed}
\email{sza9wz@virginia.edu}
 \affiliation{Department of Electrical and Computer Engineering, University of Virginia, Charlottesville, Virginia 22904, USA}
\author{Yaohua Tan}%
\affiliation{Synopsys Inc, Mountain View, California 94043, USA}

\author{Jiyuan Zheng}
\affiliation{Beijing National Research Center for Information Science and Technology, Tsinghua University, 100084, Beijing, China}%

\author{Joe C. Campbell}
\affiliation{Department of Electrical and Computer Engineering, University of Virginia, Charlottesville, Virginia 22904, USA}

\author{Avik W. Ghosh}
\affiliation{Department of Electrical and Computer Engineering, University of Virginia, Charlottesville, Virginia 22904, USA}
\affiliation{Department of Physics, University of Virginia, Charlottesville, Virginia 22904, USA}

\date{\today}

\begin{abstract}
A series of III-V ternary and quarternary digital alloy avalanche photodiodes (APDs) have recently been seen to exhibit very low excess noise. Using band inversion of an environment-dependent atomistic tight binding description of short period superlattices, we argue that a combination of increased effective mass, minigaps and band split-off are primarily responsible for the observed superior performance. These properties significantly limit the ionization rate of one carrier type, either holes or electrons, making the avalanche multiplication process unipolar in nature. The unipolar behavior in turn reduces the stochasticity of the multiplication gain. The effects of  band folding on carrier transport are studied using the Non-Equilibrium Green's Function Method that accounts for quantum tunneling, and Boltzmann Transport Equation model for scattering. It is shown here that carrier transport by intraband tunneling and optical phonon scattering are reduced in materials with low excess noise. Based on our calculations, we propose five simple inequalities that can be used to approximately evaluate the suitability of digital alloys for designing low noise photodetectors. We evaluate the performance of multiple digital alloys using these criteria and demonstrate their validity.
\end{abstract}

\maketitle

\section{\label{Introduction}Introduction}
The demand for efficient optical detectors is constantly growing due to rapid developments in telecommunication, light imaging, detection and ranging (LIDAR) systems and other military and research fields \cite{Tosi, CAMPBELL2008221, bertone2007avalanche, Mitra2006, Williams2017, Nada2020, Pasquinelli2020}. Photodetectors are increasingly being incorporated in photonic integrated circuits for Internet of Things and 5G communications \cite{li20185g,Chowdhury5G,Liu_III_V}. These applications  require higher sensitivity in comparison to traditional \textit{p-i-n} photodiodes \cite{apd_recent}. Avalanche photodiodes (APDs) are often deployed instead due to their higher sensitivity, enabled by their intrinsic gain mechanism. However,  the stochastic nature of the impact ionization process of APDs adds an excess noise factor $F(M)= \langle m^2\rangle/\langle m\rangle^2 = kM+(1-k)(2-1/M)$ to the shot noise current, $\langle i_{shot}^{2}\rangle =2qIM^{2} F(M) \Delta f$ \cite{mcintyre1966multiplication,Teich1986,Teich1990}. Here, $q$ is the electron charge, $I$ is the total photo plus dark current, $m$ is the per primary electron avalanche gain, $M = \langle m\rangle$ the average multiplication gain and $\Delta f$ is the bandwidth. A low value of $k$, which is the ratio of hole ionization coefficient $\beta$ to the electron ionization coefficient, $\alpha$, is desirable for designing low-noise n-type APDs. This ratio stipulates that for pure electron injection, a significantly lower hole ionization than the electron ionization rate leads to reduced shot noise. If impact ionization is caused by pure hole injection, $k$ in the equation will be replaced by $1/k$. 

Recently, several III-V digital alloys, i.e., short-period superlattices with binary components stacked alternately in a periodic manner, were found to exhibit extremely low noise currents and a high gain-bandwidth product in the short-infrared wavelength spectrum \cite{InAlAs_expt,AlInAsSb_expt,AlAsSb_expt}. Characterization of InAlAs, AlInAsSb and AlAsSb digital alloy APDs have shown very small values of $k$, \cite{InAlAs_expt,AlInAsSb_expt,AlAsSb_expt} whereas other digital alloys, like InGaAs and AlGaAs, demonstrate much higher $k$ value \cite{InGaAs_expt,AlGaAs_expt}. Based on previous full-band Monte Carlo simulations, \cite{AlInAsSb_MC, InAlAs_MC,InAlAs_MC2} the low $k$ has been attributed to the presence of superlattice minigaps inside the valence band of the material bandstructure, along with an enhanced effective mass arising from the lower band-width available to the holes. Such valence band minigaps often co-exist with similar (but not symmetrical) minigaps in the conduction band. However, electrons in the conduction band typically have very low effective mass, which allows quantum tunneling and enhanced phonon scattering to circumvent minigaps in the conduction band. Furthermore certain digital alloys showing mini-gaps do not exhibit low noise, and the reason behind that has not yet been addressed. Our postulate is that a combination of valence band minigap, a large separation between tight-hole and split-off bands, and corresponding enhanced hole effective mass tend to limit hole ionization coefficient. A comprehensive analysis is clearly necessary to understand the carrier impact ionization in these materials. 

In this paper, we employ a fully atomistic, Environment-Dependent Tight Bindng (EDTB) model, \cite{TanETB} calibrated to Density Functional Theory (DFT) bandstructure as well as wavefunctions, to compute the bandstructures of several III-V digital alloys. Using a full three-dimensional quantum kinetic Non-Equilibrium Green's Method (NEGF) formalism with the EDTB Hamiltonian as input, we compute the ballistic transmission across these digital alloys that accounts for intraband quantum tunneling across minigaps and light-hole/split-off bands offset. Additionally, a full-band Boltzmann transport solver is employed to determine the energy resolved carrier occupation probability under the influence of an electric field in order to study the effect of optical phonon scattering in these short-period superlattices. The calculations are performed using computational resources at University of Virginia and XSEDE \cite{xsede}. Using these transport formalisms, we elucidate the impact of minigap sizes, light-hole/split-off band offset and effective masses on carrier transport in the valence band.  

Our simulations demonstrate that the squashing of subbands into tighter band-widths, such as arising from minigap formation, or the engineering of large light-hole/split-off band offset lead to the suppression in transport of one carrier type, by resisting quantum tunneling or phonon-assisted thermal jumps. For InAlAs, the improved performance is primarily due to the minigaps generated by the digital alloy periodicity and the corresponding enhanced effective mass. For AlInAsSb and AlAsSb, the gain is a combination of minigaps, large effective mass and LH/SO offset. The LH/SO offsets in these  two alloys results arise from the strong spin-orbit coupling due to the Sb atoms, a characteristic which is also observed in their random alloy counterparts that exhibit low noise. A quantitative comparison of the various alloy gains measured is presented in the last two columns of Table IV.

The unique superlattice structure of the digital alloys opens the possibility for designing new low-noise alloy combinations for detection of other frequency ranges. Ideally, it is easier and cheaper to at first computationally study the suitability of the alloys for achieving low noise before actually fabricating these. For this purpose, we need a set of design criteria for judging the alloy performance using theoretically calculated parameters. Based on our simulations, we propose five simple inequalities that can be used to judge the suitability of digital alloys for use in low-noise APDs. We judge the aptness of five existing digital alloys- InAlAs, InGaAs, AlGaAs, AlInAsSb and AlAsSb. We observe that the inequalities provide a good benchmark for gauging the applicability of digital alloys for use in low-noise APDs.

\section{\label{Simulation}Simulation Method}
\subsection{Environment Dependent Tight Binding and Band Unfolding for atomistic description}\label{bandstructure_sec}

\begin{figure}[b]
\includegraphics[width=0.45\textwidth]{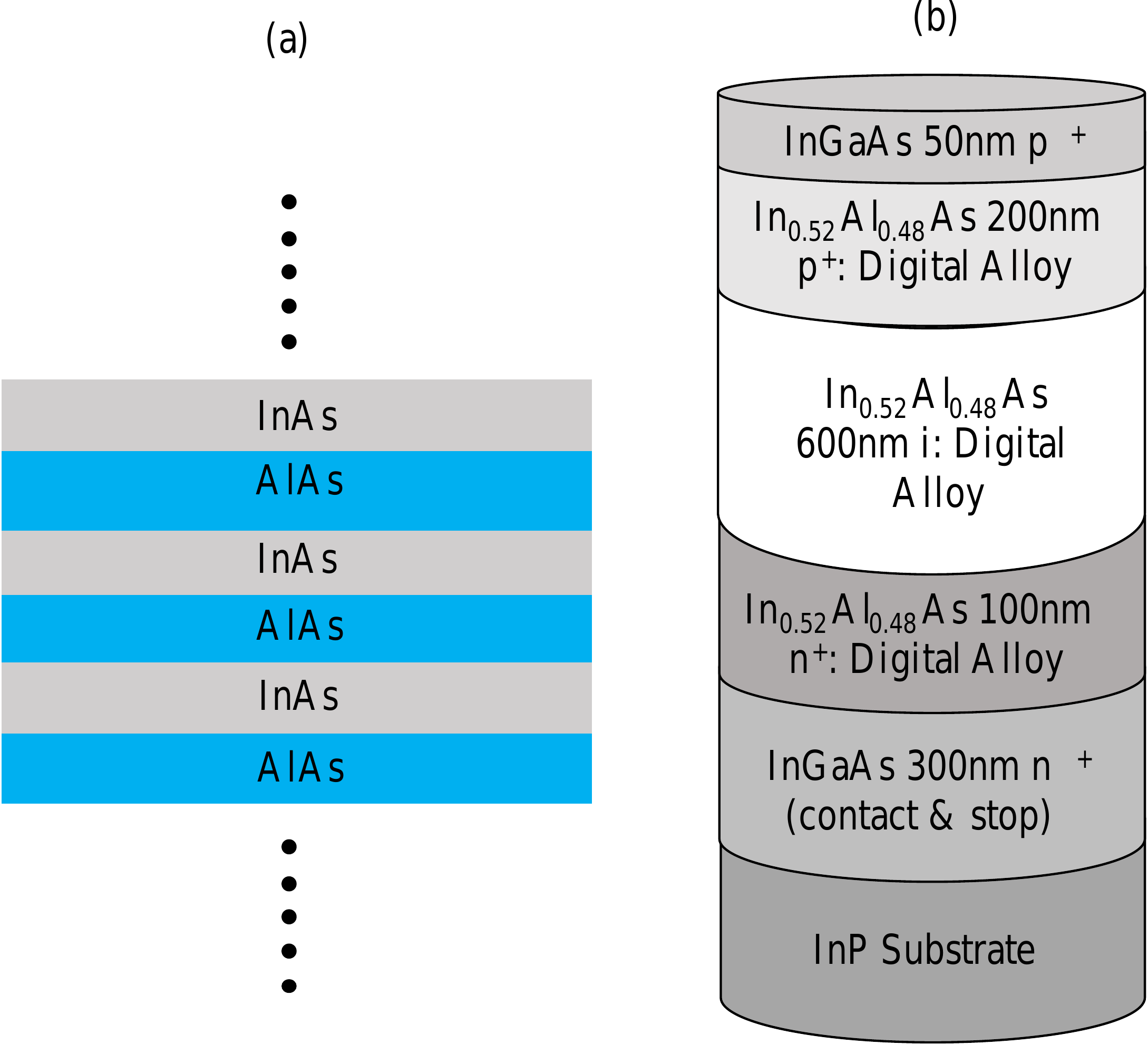}
\caption{\label{structure} (a) Digital alloy structure (b) typical structure of an APD
}
\end{figure}

In order to understand the influence of minigap filtering in digital alloy structures, an accurate band structure over the entire Brillouin zone is required. The periodic structure of the InAlAs digital alloy is shown in Fig.~\ref{structure}(a) and Fig.~\ref{structure}(b) shows the typical structure of a  \textit{p-i-n} APD. We have developed an Environment-Dependent Tight Binding (EDTB) Model to accurately calculate the band structure of alloys \cite{TanETB,TanSi}. Traditional tight binding models are calibrated directly to bulk bandstructures near their high symmmetry points and not to the underlying chemical orbital basis sets \cite{TanSi}. These models are not easily transferable to significantly strained surfaces and interfaces where the environment has a significant impact on their material chemistry. In other words, the tight binding parameters work directly with the eigenvalues (E-k) and not with the full eigenvectors. While the crystallographic point group symmetry is enforced by the angular transformations of the orbitals, the radial components of the Bloch wavefunctions, which determine bonding and tunneling properties, are left uncalibrated. Previously, in order to incorporate accuracy of radial components, an Extended H\"uckel theory \cite{huckel_cnt,huckel_silicon} was used that incorporated explicit Wannier basis sets created from non-orthogonal atomic orbitals that were fitted to Density Functional Theory for the bulk Hamiltonian. The  fitted basis sets were transferrable to other environments by simply recomputing the orbital matrix elements that the bonding terms were assumed to be proportional to. As an alternative, the EDTB model employs conventional orthogonal Wannier like basis sets. The tight binding parameters of this model are generated by fitting to both Hybrid functional (HSE06) \cite{heyd2003hybrid} band structures and orbital resolved wave functions. Our tight binding model can incorporate strain and interface induced changes in the environment by tracking changes in the neighboring atomic coordinates, bond lengths and bond angles. The onsite elements of each atom have contributions from all its neighboring atoms.  The fitting targets include unstrained and strained bulk III-V materials as well as select alloys. We have shown in the past that our tight binding model has the capability of matching the hybrid functional band structures for bulk, strained layers and superlattices \cite{TanETB,AhmedTFET}.

The band structures of the alloys contain a massive number of spaghetti-like bands due to the large supercell of the system that translates to a small Brillouin zone with closely separated minibands and minigaps. In order to transform the complicated band structure into something tractable, we employ the technique of band unfolding \cite{tan_unfolding,boykin_unfolding1,boykin_unfolding2}. This method involves projecting the eigenvalues back to the extended Brillouin zone of the primitive unit cell of either component, with weights set by decomposing individual eigenfunctions into multiple Bloch wavefunctions with different wave vectors in the Brillouin zone of the original primitive unit cell. The supercell eigenvector $\mid  \vec{K}m \rangle$ is expressible in terms of the linear combination of primitive eigenvectors $\mid  \vec{k_i}n \rangle$. The eigenstate $E_p$ of an atom with wavector $k$ can be expressed as a linear combination of atomic-orbital wavefunctions. The supercell electron wavefunction $| \psi_{m\vec{K}}^{SC} \rangle$ can be written as a linear combination of electron wavefunctions in the primitive cell as \cite{InAlAs_expt}

\begin{equation}
| \psi_{m\vec{K}}^{SC} \rangle = \sum_n a\left(\vec{k_i},n;\vec{K},m\right) |\psi_{n\vec{k_i}}^{PC} \rangle 
\end{equation}

\begin{equation}
\vec{k_i} \in \{ \vec{\tilde{k_i}}\}\nonumber 
\end{equation}
where, $| \psi_{n\vec{k_i}}^{PC} \rangle$ is the electron wavefunction for the wave vector $\vec{k_i}$ in the $n$th band of the primitive cell. Here, $\vec{\bm{K}}$ and $\vec{\bm{k}}$ denote the reciprocal vector in supercell and primitive cell respectively. The folding vector $\vec{\bm{G}}_{\vec{\bm{k}}\rightarrow \vec{\bm{K}}}$  contains the projection relationship and is expressed as

\begin{equation}
    \vec{\bm{K}}=\vec{\bm{k}}-\vec{\bm{G}}_{\vec{\bm{k}}\rightarrow \vec{\bm{K}}} ~.
\end{equation}
The projection of the supercell wavefunction $| \psi_{m\vec{K}}^{SC} \rangle$ into the primitive cell wavefunction $| \psi_{n\vec{k_i}}^{PC} \rangle$ is given as

\begin{equation}
 P_{m \vec{K}}=\sum_n \mid \langle \psi_{m \vec{K}}^{SC} | \psi_{n \vec{k_i}}^{PC}\rangle \mid^2     ~.    
\end{equation}
Plotting these projection coefficients gives a cleaner picture of the band evolution from the individual primitive components to the superlattice bands.

\subsection{Non-Equilibrium Green's Function Method for coherent transmission} \label{NEGF}
Under the influence of a large electric field it is possible for carriers to move across minigaps by means of quantum tunneling. Such transport involves a sum of complex transmissions limited by wavefunction symmetry between several minibands.
We make use of the Non-Equilibrium Green's Function formalism to compute the ballistic transmission and study the influence of minigaps on quantum tunneling in digital alloys. The digital alloys we are interested in studying are translationally invariant in the plane perpendicular to the growth direction and have finite non-periodic hopping in the transport (growth) direction. Thus, we need a device Hamiltonian $H$ whose basis is Fourier transformed into $k$-space in the perpendicular $x-y$ plane but is in real space in the $z$ growth direction, i.e., $H\left(r_z,k_x,k_y\right)$. Conventionally, this can be done with a DFT Hamiltonian in real space, $H\left(r_z,r_x,r_y\right)$, which is Fourier transformed along the transverse axes to get $H\left(r_z,k_x,k_y\right)$. However, DFT Hamiltonians are complex and sometimes do not match with bulk material bandstructure. Thus, it is simpler to utilize a tight binding Hamiltonian whose $E-\vec{k}$s are calibrated to bulk bandstructure, and inverse transform along the growth direction. 

The matrix elements of the 3D EDTB Hamiltonian are given in the basis of symmetrically orthogonalized atomic orbitals $\left|nb\textbf{R}\right>$. Here $\textbf{R}$ denotes the position of the atom, $n$ is the orbital type ($s,p,d$ or $s^*$) and $b$ denotes the type of atom (cation or anion). The Hamiltonian can also be represented in $k-$space basis $\left|nb\textbf{k}\right>$ by Fourier transforming the elements of the real-space Hamiltonian. The 3D Hamiltonian is then converted into a quasi-1D Hamiltonian \cite{stovneng1993multiband}. The Hamiltonian elements can be represented in the basis $\left|nbj\textbf{k}_{||}\right>$ with``parallel'' momentum $\textbf{k}_{||}=(k_x,k_y)$ and ``perpendicular'' position $x_j=a_L/4$ as parameters. For a zinc-blende crystal, the distance between nearest-neighbour planes is one-fourth the lattice constant $a_L$. The 3D Hamiltonian is converted to the the quasi-1D one by means of a partial Fourier transform \cite{stovneng1993multiband,stickler2013theory}: 
\begin{equation}
   \left|nbj\textbf{k}_{||}\right>=L_{BZ}^{-1/2}\int dk_z e^{-ik_z ja_L/4} \left|nb\textbf{k}\right> ~.
\end{equation}
Here $L_{BZ} =8\pi/a_L$ is the length
of the one-dimensional (1D) Brillouin zone over which
the $k_z$ integral is taken. The quasi-1D Hamiltonian is position dependent in the growth direction. Thus, we are able to utilize the accurate bandstructure capibility of the EDTB. 

In presence of contacts, the time-independent open boundary Schr\"odinger equation reads 
\begin{equation}
    (EI-H-\Sigma_1-\Sigma_2)\Psi = S_1+S_2
\end{equation}where, $E$ represents energy, $I$ denotes identity matrix and $\Sigma_{1,2}$ are the self-energies for the left and right contact respectively describing electron outflow, while $S_{1,2}$ are the inflow wavefunctions. The solution to this equation is $\Psi = G(S_1+S_2)$, where the Green's function \cite{datta2000nanoscale}
\begin{equation}
    G(E)=\left[EI-H-\Sigma_1-\Sigma_2\right]^{-1} ~.
\end{equation}
Here $H$ includes the applied potential, added to the onsite 1D elements. Assuming the contacts are held in local equilibria with bias-separated quasifermi levels $\mu_{1,2}$, we can write the bilinear thermal average $\langle S_iS^\dagger_i \rangle = \Gamma_if(E-\mu_i)$ where $f$ is the Fermi-Dirac distribution and $\Gamma_{1,2} = i(\Sigma_{1,2}-\Sigma_{1,2}^\dagger)$ denoting the broadening matrices of the two contacts. The equal time current $I = q(d/dt + d/dt^\prime)Tr\langle \Psi^\dagger(t)\Psi(t^\prime)\rangle|_{t=t^\prime}$ then takes the Landauer form $I = (q/h)\int dET(f_1-f_2)$, where the 
coherent transmission between the two contacts is set by the Fisher-Lee formula 
\begin{equation}
    T(E)=Tr\left[\Gamma_1 G \Gamma_2 G^\dagger \right]
\end{equation}
where $Tr$ represents the trace operator. 
The energy resolved net current density from the layer $m$ to layer $m+1$ is expressed as\cite{stovneng1993multiband}: 

\begin{eqnarray}
    J_{m,m+1}(E) &=& -\frac{iq}{h} \int \frac{\textbf{k}_{||}}{(2\pi)^2}  Tr[G^{n,p}_{m+1,m}H_{m,m+1} \\
    & & -G^{n,p}_{m,m+1}H_{m+1,m}] \nonumber
\end{eqnarray}
where, $G^{n} = \langle \psi^\dagger\psi\rangle$ and $G^p = \langle \psi\psi^\dagger\rangle$  represent electron ($n$) and hole density ($p$) correspondingly and $H_{m,m+1}$ is the tight binding hopping element between layers $m$ and $m+1$ along the transport/growth direction.

\subsection{Boltzmann Transport Model for incoherent scattering} \label{BTE}
The NEGF approach is particularly suited to ballistic transport where coherent quantum effects dominate. Incoherent scattering requires a self-consistent Born approximation which is computationally quite involved. We need a practical treatment of scattering. 
Under an external electric field, the carrier distributions in digital alloys no longer follow a local Fermi-distribution, but re-distribute over real-space and momentum space. To understand the carrier distribution under electric field in digital alloys, we employed the multi-band Boltzmann equation.
\begin{eqnarray}
    \vec{v}\cdot \nabla_{\textbf{r}} f_n + \vec{F}\cdot \nabla_{\textbf{k}} f_n & =& \sum_{m, \vec{p}'} S\left(\vec{p}',\vec{p}\right)f_m\left(\vec{p}'\right) \left[1-f_n\left(\vec{p}\right)\right] \\
    & & - \sum_{m,\vec{p}'} S\left(\vec{p},\vec{p}'\right)f_n\left(\vec{p}\right) \left[1-f_m\left(\vec{p}'\right)\right]\nonumber
\end{eqnarray}
Here, $f = f(\textbf{r},\textbf{k})$ is the carrier distribution, $n$ and $m$ are band indices, $\vec{p}$ and $\vec{p}'$ are the momenta of the carriers, and $S\left(\vec{p}',\vec{p}\right) $ is the scattering rate. The left hand side of this equation alone describes the ballistic trajectory in the phase space of carriers under electric field. 
The right hand side of the equation corresponds to the scattering processes including intra-band and inter-band scattering. 

In a homogenenous system where the electric field is a constant, the distribution function is independent of position,  $\nabla_{\textbf{r}} f = 0$ and the equation is reduced to
\begin{eqnarray}\label{bte_kspace}
    \vec{F}\cdot \nabla_{\textbf{k}} f_n & =& \sum_{m,\vec{p}'} S\left(\vec{p}',\vec{p}\right)f_m\left(\vec{p}'\right) \left[1-f_n\left(\vec{p}\right)\right] \\
    & & - \sum_{m,\vec{p}'} S\left(\vec{p},\vec{p}'\right)f_n\left(\vec{p}\right) \left[1-f_m\left(\vec{p}'\right)\right]\nonumber ~.
\end{eqnarray}
For APDs, it is critical to consider  optical phonon scattering, which is the dominant process besides tunneling that allows  carriers to overcome the minigap arising in the band structures of digital alloys. The optical phonon has a non-trivial energy of $\hbar \omega_{opt}$ that can be absorbed or emitted by carriers. The scattering rate $S\left(\vec{p}',\vec{p}\right) $ has the form set by Fermi's Golden Rule
\begin{equation}
    S\left(\vec{p}',\vec{p}\right) = \frac{2\pi}{\hbar} \left|H_{\vec{p}, \vec{p}'}\right|^2 \delta_{\vec{p}',\vec{p}\pm \vec{\beta}} \delta \left(E(\vec{p}')-E(\vec{p})\pm \hbar \omega_{opt}\right) ~.
\end{equation}
The $E(\vec{p})$ and $E(\vec{p}')$ are band structures of digital alloy calculated by the tight binding model. $H_{p,p^\prime}$ can be calculated by evaluating electron-phonon coupling matrix elements explicitly.
In this work, we extract a constant effective constant scattering strength $H_{\vec{p}, \vec{p}'}$ from experimental mobility $\mu$. The scattering lifetime $\tau$, which is $1/S\left(\vec{p}',\vec{p}\right)$, can be extracted from the mobility using $\mu=q\tau/m^*$. Due to lack of experimental mobilities of the digital alloys, we considered the average of the binary constituent room temperature mobilities for extracting the lifetime. A simple average is done since the binary constituents in periods of most of the digital alloys considered here are equally divided. In using room temperature values the underlying assumption is that the dominant scattering mechanism here is phonon scattering due to large phonon population. Ionized impurity scattering is considered to be much lower due to digital alloys having clean interfaces \cite{AlInAsSb_expt}. It is then possible to extract $H_{\vec{p}, \vec{p}'}$ from the scattering lifetime. To get the equilibrium solution, we solve Eq.~\ref{bte_kspace} self-consistently, starting from an initial distribution $f = \delta_{\vec{k},0}$.

A detailed model of carrier transport in APDs also requires a NEGF treatment of impact ionization self-energies and a Blanter-Buttiker approach to extract shot noise, but we leave that to future work. Our focus here is on conductive near-ballistic transport, and the role of quantum tunneling and perturbative phonon scattering in circumventing this.

\section{\label{results}Results and Discussion}

\begin{figure}[t]
\includegraphics[width=0.48\textwidth]{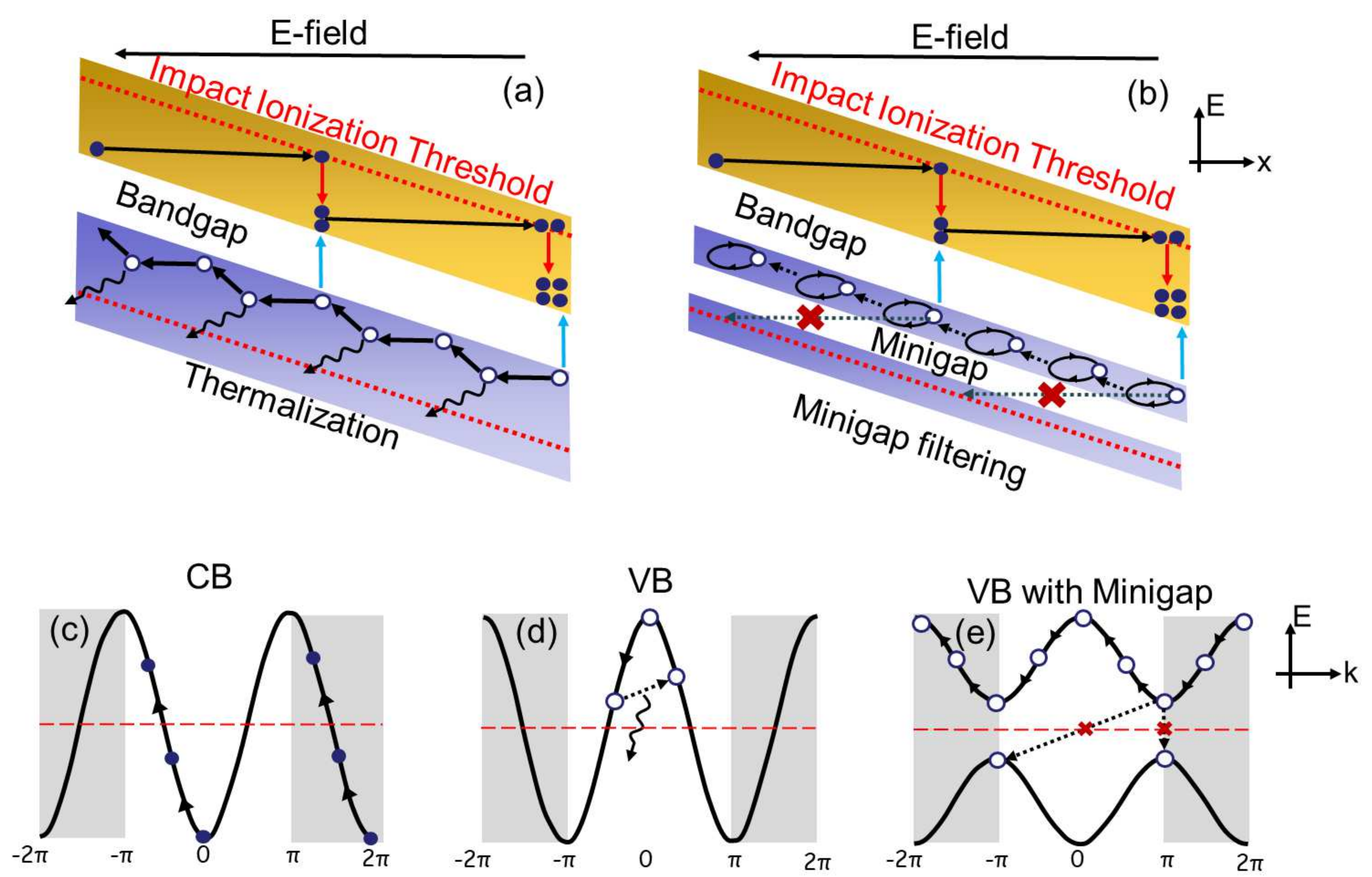}
\caption{\label{fig:impact_ionization_mechnism} Impact ionization process in normal (random alloy) APD and superlattice APD.
In both APDs, it is easier for electrons to gain energy and reach the impact ionization threshold (c). In normal APDs (a), holes find it harder to gain high energy compared to electrons because of thermalization. The hole energy is reduced by thermalization due to various scattering processes as shown in (d).In superlattice APD (b), the existence of minigaps makes it harder for holes to reach higher energies. The minigaps acts as barrier that prevent holes from moving to the lower valence bands. In the plots, the y-axis $E$ is the total energy (kinetic+potential) meaning in between inelastic scattering events the particles travel horizontally.
}
\end{figure}

There are three common ways to achieve low noise and high gain-bandwidth product - selecting a semiconductor with favorable impact ionization coefficients, scaling the multiplication region to exploit the non-local aspect of impact ionization, and impact ionization engineering using appropriately designed heterojunctions \cite{apd_recent}. Typically, the lower hole impact ionization coefficient in semiconductors is due to stronger scattering in the valence bands, as depicted in Fig.~\ref{fig:impact_ionization_mechnism}(a). Previously, the lowest noise with favorable impact ionization characteristics were realized with Si in the visible and near-infrared range, \cite{lee1964ionization,conradi1972distribution,grant1973electron,kaneda1976model} and InAs \cite{marshall2008electron,marshall2011high,sun2014record,sun2012high,ker2012inas} and HgCdTe \cite{beck2001mwir,beck2004hgcdte} in the mid-infrared spectrum. In comparison, InGaAs/InAlAs \cite{InGaAs_random,InAlAs_random} random alloy APDs exhibit significantly higher noise than Si, HgCdTe or InAs, which are the highest performance telecommunications APDs. In the recent past, digital alloy InAlAs APDs have demonstrated lower noise compared to their random alloy counterpart \cite{InAlAs_expt}. This seems a surprise, as the suppression of one carrier type (the opposite of ballistic flow expected in an ordered structure) is necessary for low excess noise. Initially, the low value of $k$ in InAlAs was attributed to the presence of minigaps \cite{InAlAs_MC2}. However, minigaps were also observed in InGaAs digital alloy APDs which have higher excess noise\cite{ahmed2018apd,InGaAs_expt}. So, a clearer understanding of the minigap physics was needed and hence a comprehensive study was required.

\begin{figure}[t]
\includegraphics[width=0.45\textwidth]{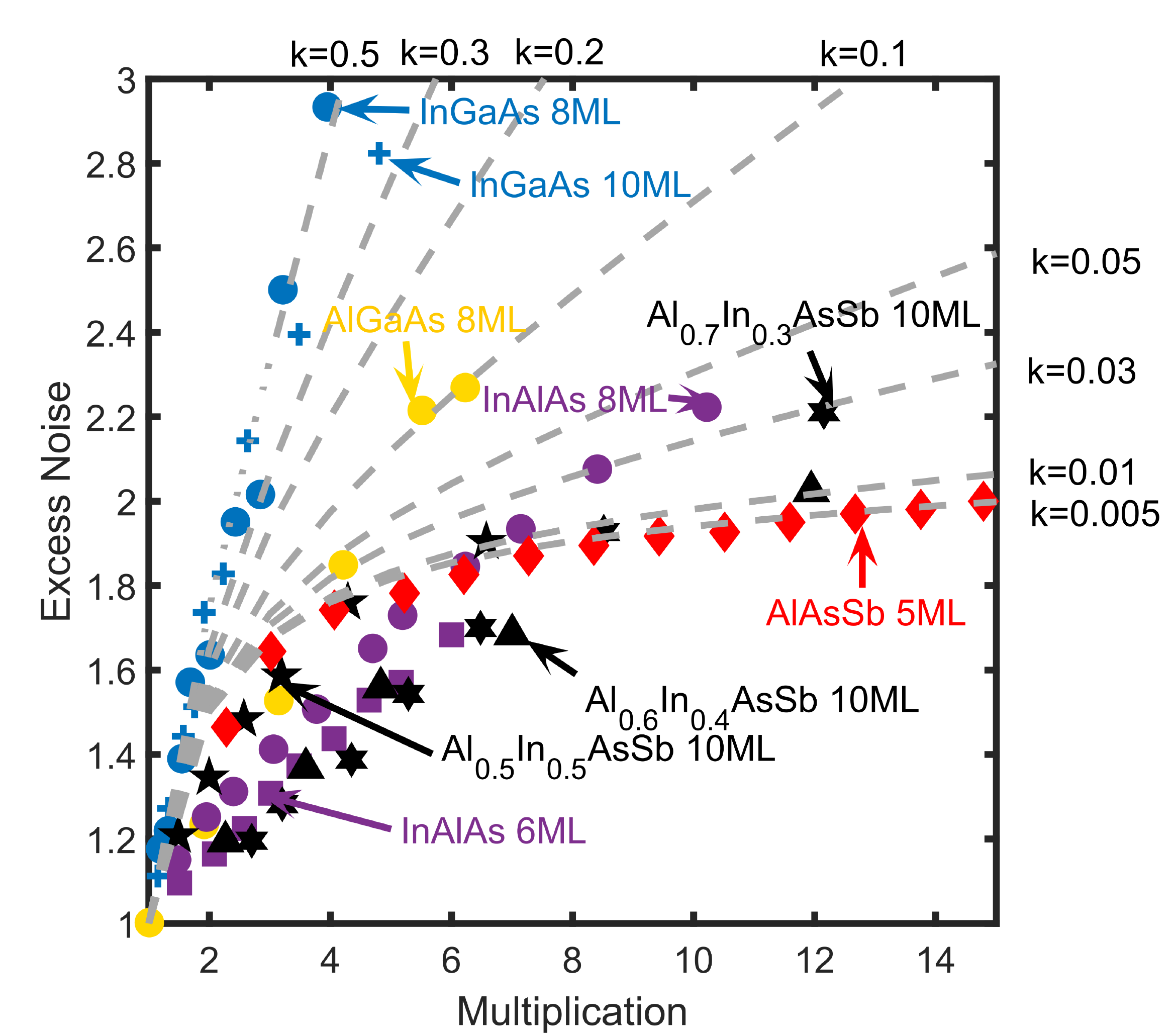}
\caption{\label{fig:expt k values} Experimentally measured Excess noise vs. Multiplication gain of InGaAs, AlGaAs, InAlAs, AlInAsSb and AlAsSb digital alloys are shown here \cite{InAlAs_expt,InGaAs_expt,AlGaAs_expt,AlInAsSb_expt,AlAsSb_expt}. The dotted lines for the corresponding k’s are plotted using McIntyre’s formula \cite{mcintyre1966multiplication}.}
\end{figure}

Our recent results suggest that well defined minigaps introduced in the valence band of digital alloys suppress the density of high energy holes and thereby reduce the impact ionization greatly, as shown in Fig.~\ref{fig:impact_ionization_mechnism}(b). In a regular low-noise electron-injected APD, the electron ionization coefficient is much higher than the hole ionization coefficient. Thus, electrons can easily climb to higher kinetic energies in the conduction band, depicted in Fig.~\ref{fig:impact_ionization_mechnism}(c), and participate in the impact ionization process by gaining the impact ionization threshold energy. On the other hand holes lose energy by various inelastic scattering processes (Fig.~\ref{fig:impact_ionization_mechnism}(d)), collectively known as thermalization. Thermalization prevents holes from reaching their secondary impact ionization threshold. In superlattice APDs, minigaps provide an additional filter mechanism that prevents holes from reaching the threshold energy required to initiate secondary impact ionization. 

\begin{figure}[b]
\includegraphics[width=0.49\textwidth]{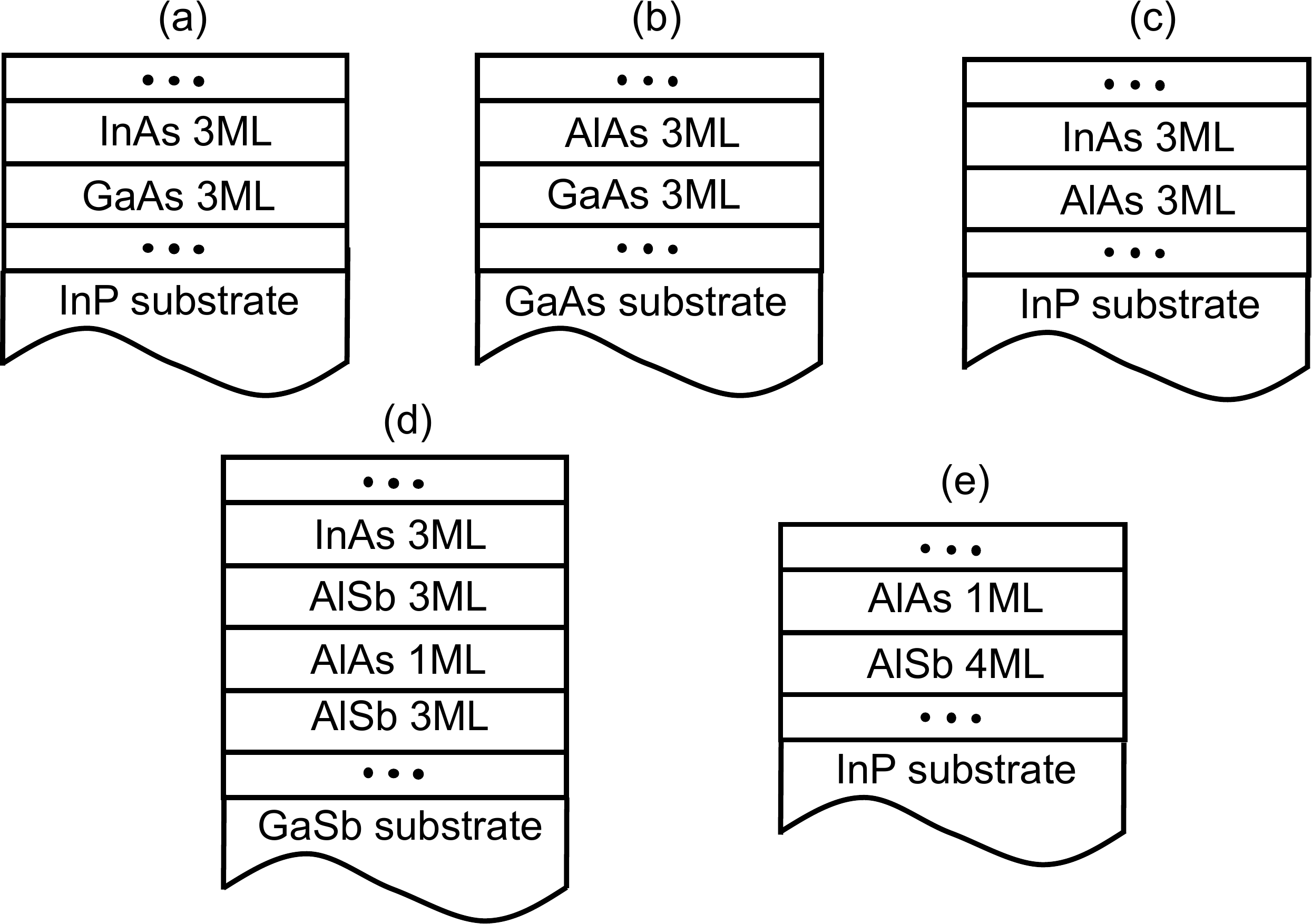}
\caption{\label{fig:material_structure} Lattice structures of (a) InGaAs, (b) AlGaAs, (c) InAlAs, (d) AlInAsSb and (e) AlAsSb digital alloys considered in this paper.}
\end{figure}

The effect of minigaps is shown in Fig.~\ref{fig:impact_ionization_mechnism}(e). However, not all digital alloy APDs exhibit low noise. 
The excess noise $F(M)$ vs. multiplication gain characteristics of experimental InGaAs, AlGaAs, InAlAs, AlInAsSb and AlAsSb digital alloy APDs are shown in Fig. \ref{fig:expt k values} \cite{InAlAs_expt,InGaAs_expt,AlGaAs_expt,AlInAsSb_expt,AlAsSb_expt}. InGaAs APDs have the highest excess noise while AlAsSb has the lowest. The dotted lines represent the theoretical $F(M)$ vs. $M$ calculated using the well known McIntyre's formula \cite{mcintyre1966multiplication}, introduced in the first paragraph of this paper. In order to understand the underlying physics in these digital alloys, an in-depth analysis of the material bandstructure and its effect on the carrier transport is required.

\begin{figure*}[t]
\includegraphics[width=1\textwidth]{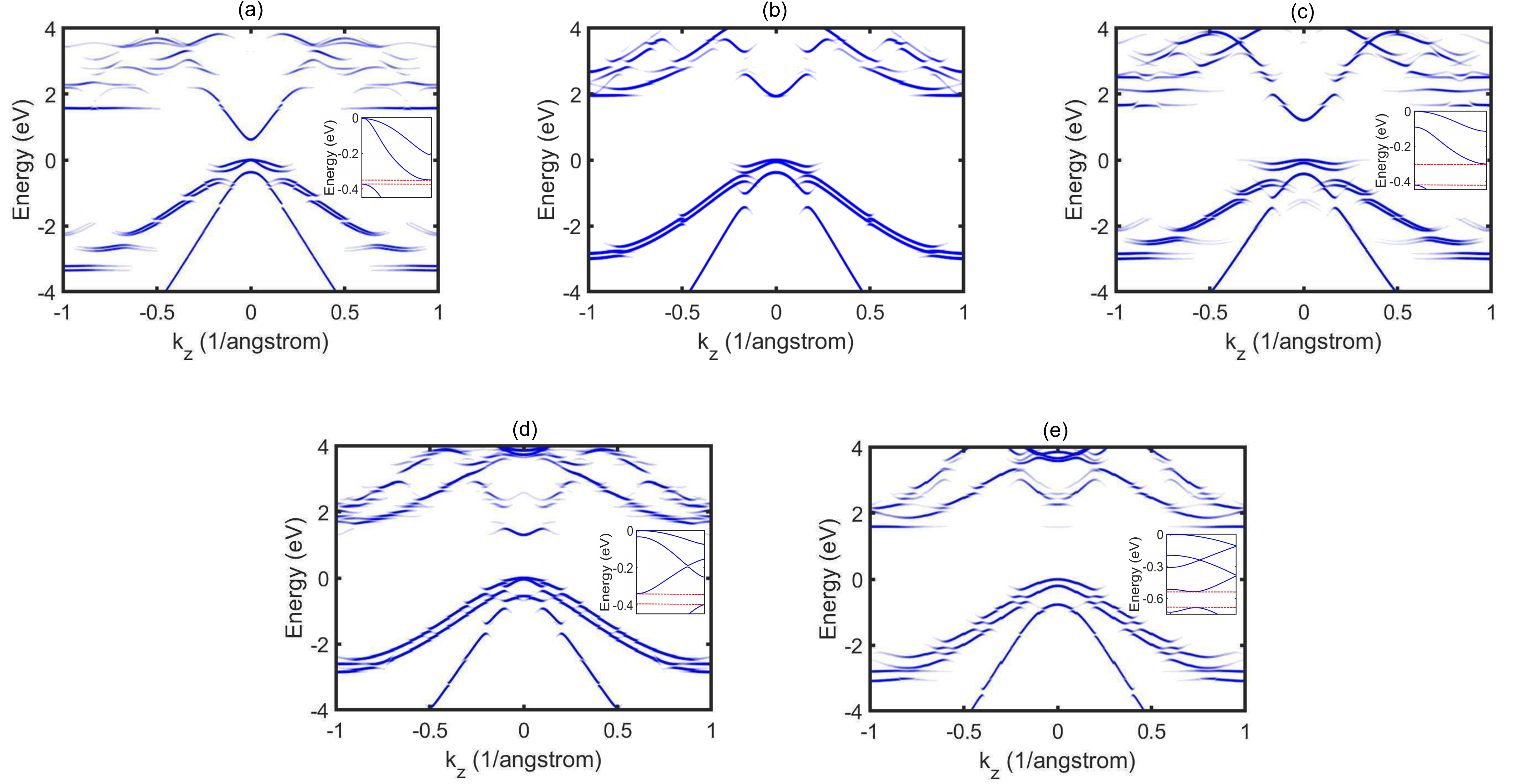}
\caption{\label{fig:bandstructure} Unfolded bandstructure of (a) 6ML InGaAs (b) 6ML AlGaAs (c) 6ML InAlAs (d) 10ML AlInAsSb  (e) 5ML AlAsSb. The minigaps of InGaAs, InAlAs, AlInAsSb and AlAsSb real bandstructures are shown in the insets.}
\end{figure*}

\begin{figure}[b]
\includegraphics[width=0.47\textwidth]{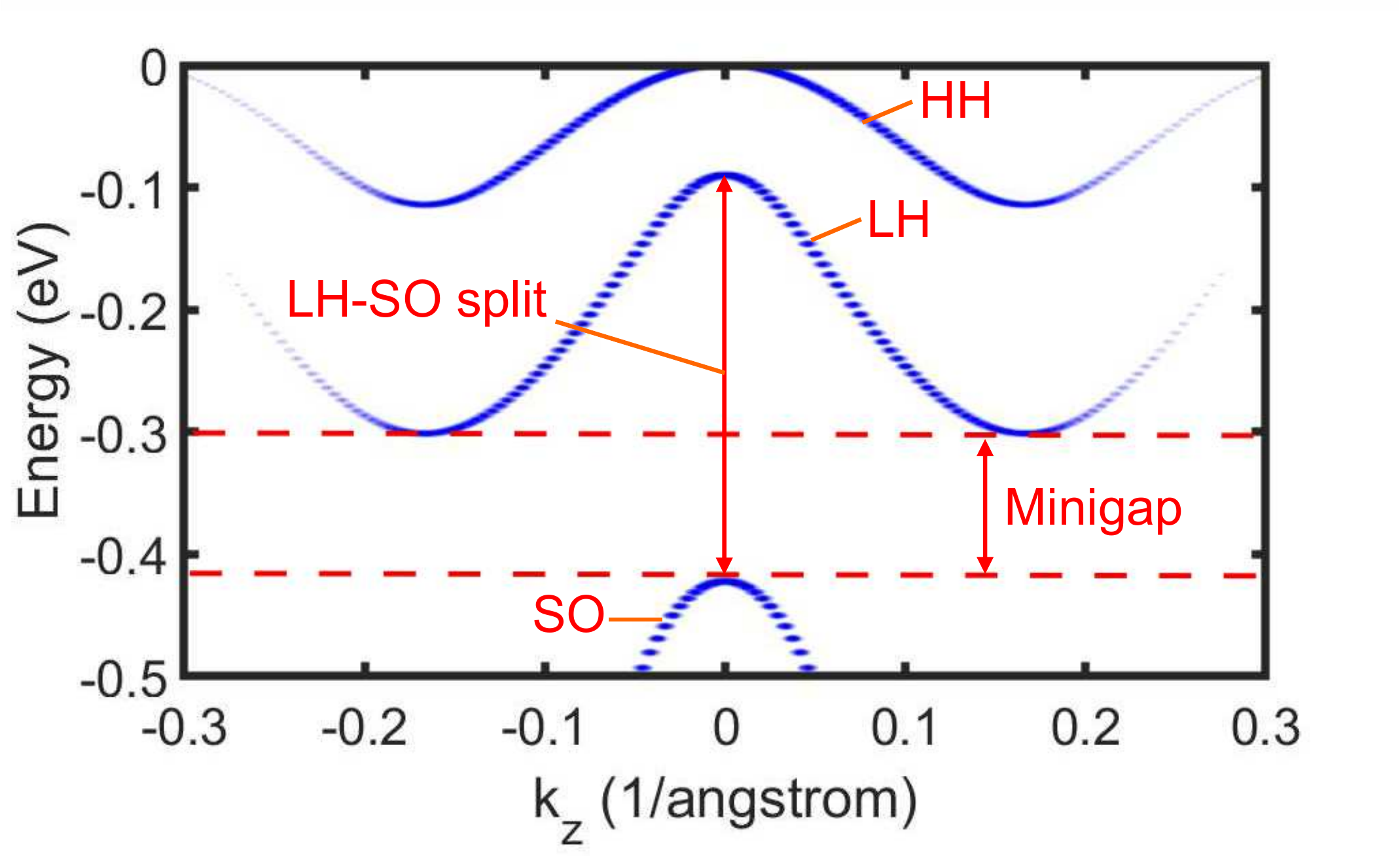}
\caption{\label{fig:InAlAs_zoomed} A magnified picture of the InAlAs valence band shows the minigap closest to the valence band edge. The split between the LH and SO at the $\Gamma$ point is also highlighted. }
\end{figure}

We calculate the atomistic DFT-calibrated EDTB bandstructure of these materials and unfold their bands using the techniques described in section \ref{bandstructure_sec}, to understand the underlying physics of their noise performance. In Fig.~\ref{fig:material_structure}, we show the periods of the different digital alloys considered- (a) 6ML InGaAs, (b) 6ML AlGaAs, (c) 6ML InAlAs, (d) 10ML Al$_{0.7}$In$_{0.3}$AsSb and (e) 5ML AlAsSb. Here, 6ML InGaAs includes 3ML InAs and 3ML GaAs, 6ML AlGaAs has 3ML AlAs and 3ML GaAs, and 6 ML InAlAs has 3ML InAs and 3ML AlAs. 10ML Al$_{0.7}$In$_{0.3}$AsSb consists of 3ML AlSb, 1ML AlAs, 3ML AlAs and 3ML InAs in its period. AlAsSb has 4ML AlSb and 1ML AlAs. 
The unfolded bandstructures of these alloys are shown in Fig. \ref{fig:bandstructure}. We observe that minigaps exist in at least one of the valence bands (heavy-hole, light-hole or split-off) for all the material combinations. The InAlAs valence band structure is magnified in Fig. \ref{fig:InAlAs_zoomed}. The minigap between the LH and SO band is denoted in the figure. Additionally, the large separation between the LH and SO bands at the $\Gamma$ point is highlighted.

\begin{table}[b]
\centering
\begin{center}
\begin{tabular}{ |m{1.3cm}|m{0.85cm} m{0.85cm} m{0.85cm} m{0.85cm} m{0.85cm} m{0.85cm} m{0.85cm}|  }  
 \hline
 \textbf{Material} & $E_G$ (eV) & $\Delta E_{b}$ (eV) & $\Delta E_{m}$ (eV) & \textbf{HH} $m^*$ & \textbf{LH} $m^*$  & \textbf{SO} $m^*$  & $\Delta E_{LS}$ (eV)\\
 \hline
 InGaAs & 0.63 & 0.34 & 0.03 & 0.31 & 0.13 & 0.045 & 0.35 \\
 \hline
 AlGaAs &1.94 & 1.03 & 0.34 & 0.45 & 0.31 & 0.12 & 0.33 \\ \hline
 InAlAs &1.23 & 0.30 & 0.12 & 0.5 & 0.4 & 0.1 & 0.31\\  
 \hline
 AlInAsSb & 1.19 & 0.33 & 0.06 & 0.42 & 0.38 & 0.08 & 0.48\\
 \hline
 AlAsSb & 1.6 & 0.56 & 0.1 & 0.45 & 0.3 & 0.13 & 0.54 \\  
 \hline
\end{tabular}
\end{center}
\caption{Material parameters of the different digital alloys simulated in this paper.}
\label{table:1}
\end{table}

The role of the minigaps on hole localization is not identical across different alloys. For instance, the presence of minigaps in material bandstructure is not sufficient to realize low noise in APDs.
Taking a closer look at the bandstructures, we observe that the positions in energy of the minigaps with respect to the valence band edge differ from one material to another. Additionally, the minigap sizes of the different alloys vary in magnitude. A complimentary effect of the minigap size is the flattening of the energy bands, \textit{i.e.}, a large minigap size results in flatter bands around the gap. This in turn results in an increased effective mass which tends to inhibit carrier transport. Table \ref{table:1} lists the energy location of the minigap with respect to the valence band edge $\Delta E_{b}$, the minigap size $\Delta E_m$, the light-hole (LH) and split-off (SO) band effective masses and the energy difference between the LH and SO bands $\Delta E_{LS}$ at the $\Gamma$ point for the digital alloys studied.

\begin{figure}[t]
\includegraphics[width=0.45\textwidth]{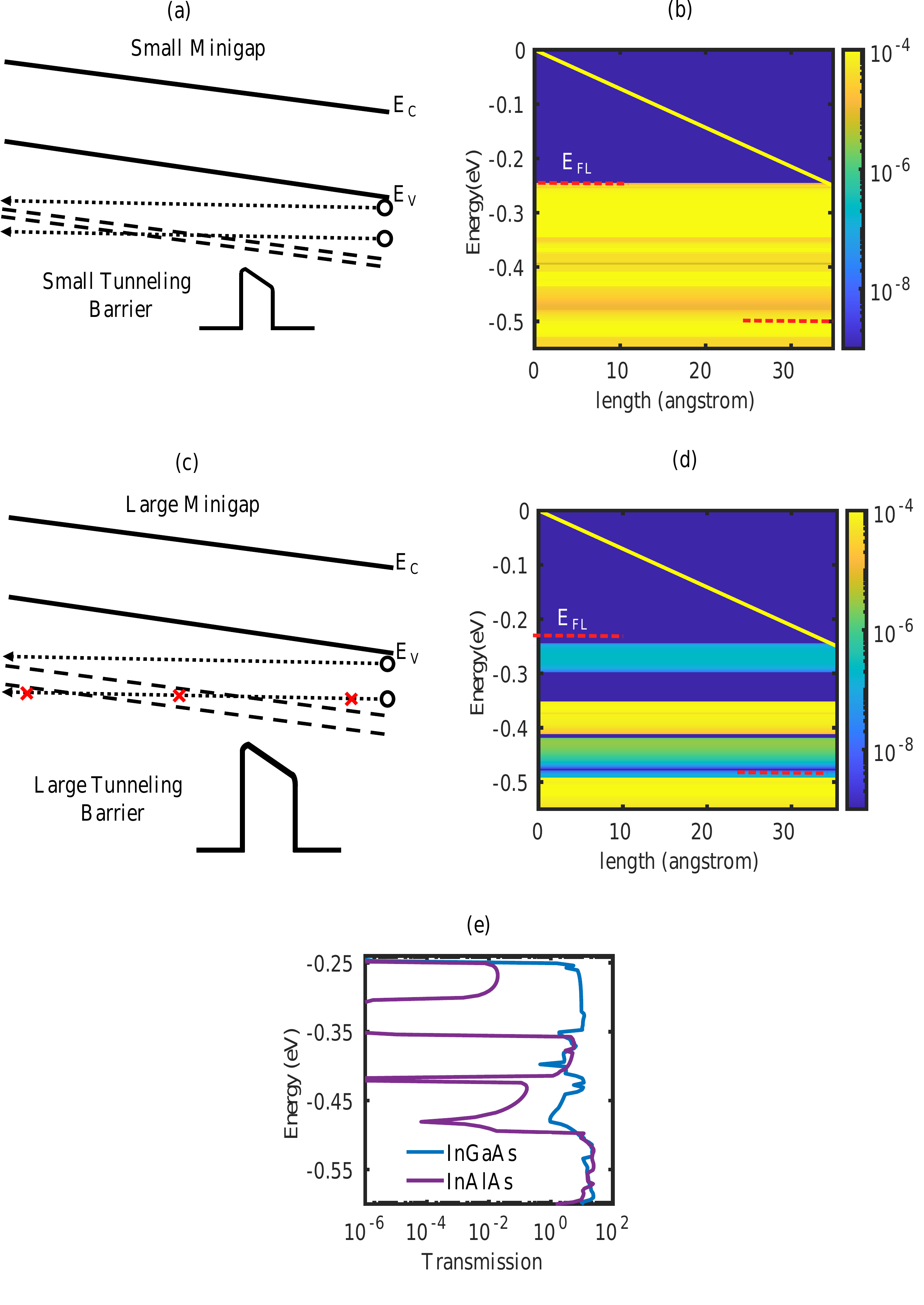}
\caption{\label{fig:minigap_current_density} Small minigaps in the valence band, as shown in (a), create a small tunneling barrier which can be overcome by holes with low mass. The spectral current density for InGaAs, which has a small minigap and small LH effective mass, is shown in (b). The current spectrum for InGaAs in the Fermi window is continuous. The creation of a large tunneling barrier by a larger minigap is shown in (c). This barrier prevents hole transmission. InAlAs has a larger minigap and LH $m^{*}$. Regions of low current density is observed within the Fermi window in the InAlAs spectral current density in (d). The large minigap in InAlAs results in reduced transmission as shown in the $T(E)$ vs. (E) plot of (e). The simulations for (b), (d) and (e) were conducted under bias of $V=0.25V$.}
\end{figure}

\begin{figure}[t]
\includegraphics[width=0.47\textwidth]{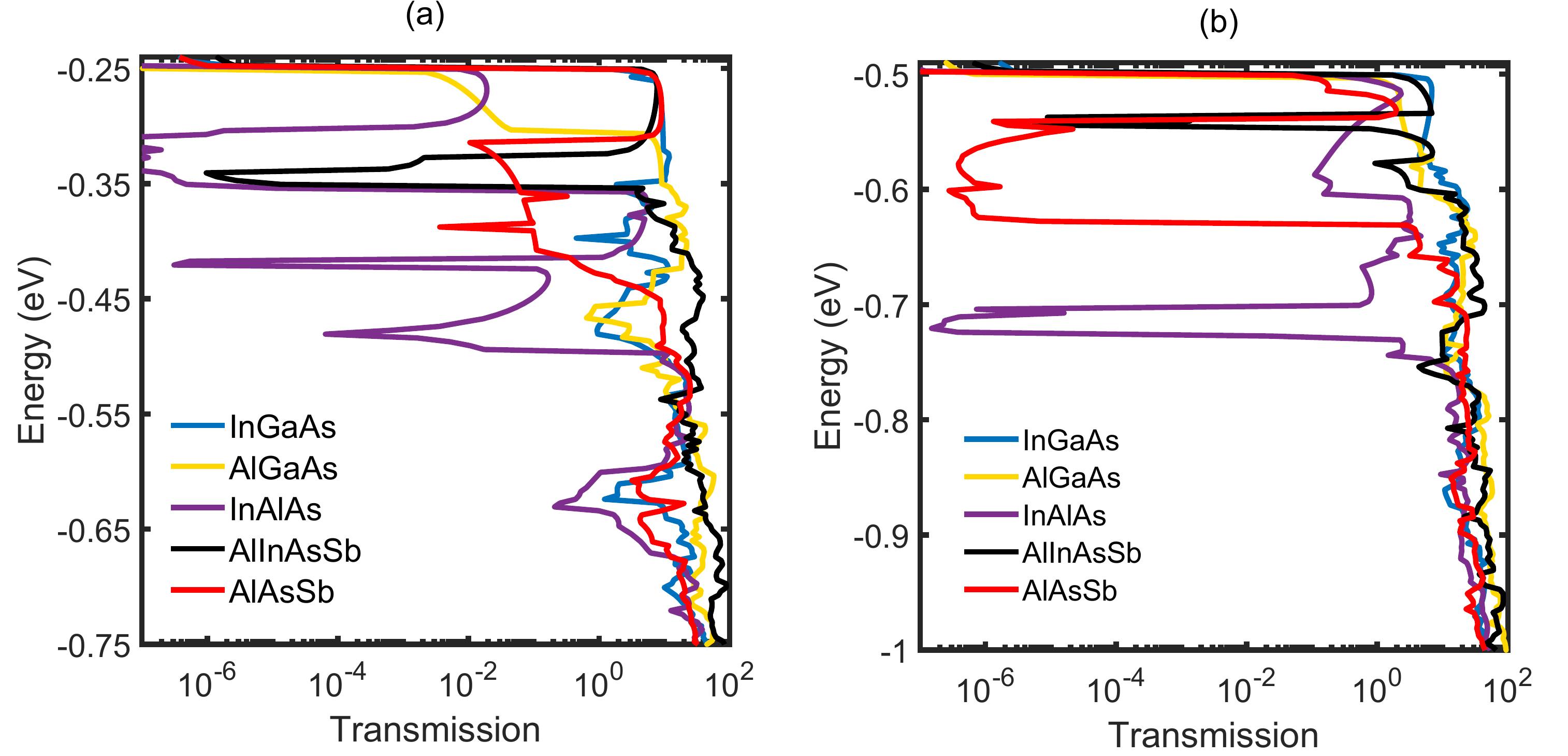}
\caption{\label{fig:negf_transmission_all} The Transmission $T(E)$ vs. Energy $E$ for all the digital alloys at $V=0.25V$ in (a) and $V=0.5V$ in (b). A $21\times 21$ grid for transverse wavevectors is used.
}
\end{figure}

We can see in the table that there are significant variations in minigap size and position between different materials. At first glance, there seems to be no direct correlation between these variations and the excess noise, prompting us to do added transport analyses.
Under high electric field, a carrier must gain at least the threshold energy, $E_{TH}$, in order to impact ionize. Typically, $E_{TH}$ is assumed to be approximately 1.5 times the material bandgap, $E_G$. Thus, in the presence of minigaps, electrons/holes must bypass these gaps by some transport mechanism in order to gain energy equivalent to $E_{TH}$. The two such major transport mechanisms are quantum mechanical tunneling and optical phonon scattering. Our transport study must incorporate these two mechanisms to understand the effectiveness of minigaps on the APD excess noise. 

We employ the NEGF formalism described in Section \ref{NEGF} to compute the ballistic transmission in the valence band as a function of energy, $T(E)$, dominated by tunneling processes. The effect of different minigap sizes is highlighted in Fig. \ref{fig:minigap_current_density}. For our simulation, we set the quasi-Fermi level of the left contact at $-qV$ below the valence band edge and quasi-Fermi level of the right contact at another $-qV$ below. This is done in order to only observe the intraband tunneling inside the valence band which is responsible for overcoming minigaps under ballistic conditions. In Fig. \ref{fig:minigap_current_density} (a), We demonstrate that a small minigap in the valence band creates a small tunneling barrier for the holes. A hole with a small enough effective mass will be able to tunnel across this barrier and render it ineffective. That is the case for InGaAs which has a LH effective mass of $0.13 m_{0}$ and $\Delta E_{m}=0.03eV$. The spectral current density for InGaAs under a bias $V=0.25V$ is shown in Fig. \ref{fig:minigap_current_density} (b). We observe that the current spectrum in the valence band is continuous in the Fermi energy window and there is no drop in transmission due to the minigap. For a large minigap, the holes encounter a larger tunneling barrier, as shown in Fig. \ref{fig:minigap_current_density} (c), preventing them from gaining the threshold energy $E_{TH}$ for secondary impact ionization. This case is operational in InAlAs digital alloys, as shown in the spectral density plot in Fig. \ref{fig:minigap_current_density} (d). InAlAs has a minigap size of $0.12eV$ and LH effective mass of $0.4m_{0}$. Within the Fermi window we see that there are regions with extremely low current due to low tunneling probability across the minigap. This is further demonstrated by the $T(E)$ vs. $E$ plot in Fig. \ref{fig:minigap_current_density} (e). Here, it is observed that there are regions of low transmission for InAlAs whereas the InGaAs transmission is continuous. This signifies that the minigaps in the InAlAs valence band are  large enough to prevent holes from gaining in kinetic energy, resulting in a low hole ionization coefficient.

In order to investigate the role of minigaps in the remaining digital alloys, we look at the transmission vs. energy plots for all the alloys. The $T(E)$ vs. $E$ characteristics for the five digital alloys are shown in Fig. \ref{fig:negf_transmission_all} for two bias conditions, (a) $V=0.25V$ and (b) $V=0.5V$. We use a $21\times21$ grid for the transverse wavevectors ($k_{x},k_{y}$) within the first Brillouin zone. For this simulation, the structure length for InGaAs, AlGaAs, InAlAs and AlAsSb is considered to be two periods. For AlInAsSb we consider one period length. This allows us to keep the structure lengths as close as possible. We consider lengths of $3.48nm$ InGaAs, $3.42nm$ AlGaAs, $3.54nm$ InAlAs, $3.06nm$ AlInAsSb and $3.08nm$ AlAsSb channels. The channel sizes chosen are small compared to actual device lengths in order to keep the computation tractable. For both the bias conditions in Fig. \ref{fig:negf_transmission_all} we see there are energy ranges for InAlAs, AlInAsSb and AlAsSb in which the transmission probability drops drastically. This low tunneling probability can be attributed to two factors. The first factor is the presence of a sizeable minigap in all directions in the material bandstructure. The other contributing factor is the separation between the LH and SO bands. This factor is partly responsible for the low transmission regions in AlInAsSb and AlAsSb, whose minigap sizes (from Table \ref{table:1}) are smaller than InAlAs but also demonstrate lower excess noise. InGaAs and AlGaAs do not have any large drop in transmission for both biases. This characteristic implies that either the minigap size is too small to affect the carrier transport like in InGaAs or there is no minigap at all as in AlGaAs.

\begin{figure}[b]
\includegraphics[width=0.47\textwidth]{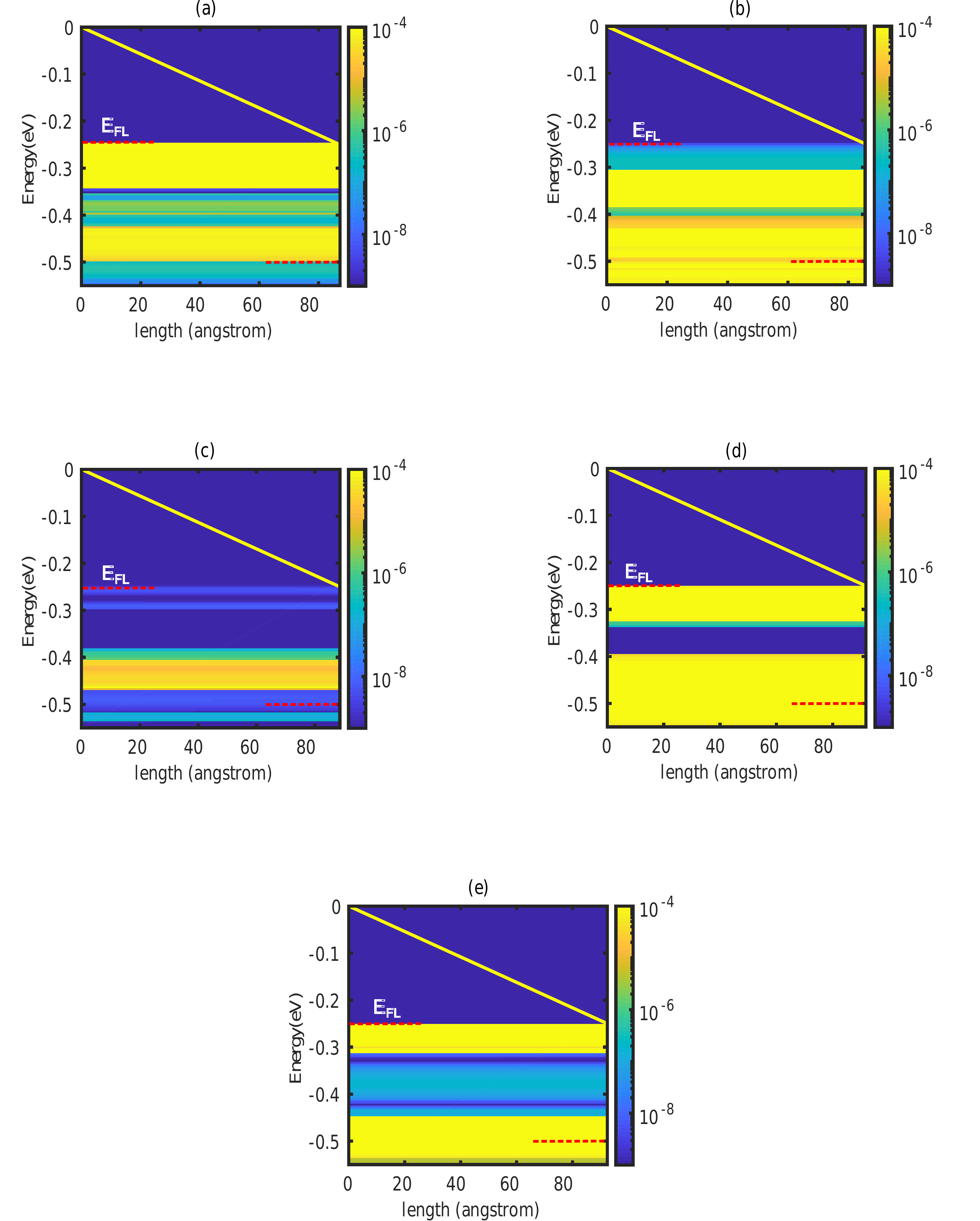}
\caption{\label{fig:current_spec_all} Energy resolved current spectral density in the valence band for (a) InGaAs, (b) AlGaAs, (c) InAlAs, (d) AlInAsSb and (e) AlAsSb. The bias for the simulation is set to $V=0.25V$ and total period length is 30 monolayers.}
\end{figure}

For further confirmation of these observations, we compute the spectral current density for the case of constant total period length of all the structures. The period size of each unit cell stays the same but the number of unit cells is increased to make the total period length the same for all alloys.  We consider the case with total period of 30MLs and voltage bias of $0.25V$. The current spectral density plots for the five digital alloys using a $15\times15$ transverse wavevector grid are shown in Fig.~\ref{fig:current_spec_all}. Smaller number of grid points are used here to save computation time. In the figure, a very small minigap is observed for InGaAs within the Fermi window and a continuous spectrum is seen for AlGaAs. Regions of low transmission/current are observed for InAlAs, AlInAsSb and AlAsSb. These observations are consistent with our previous calculations. We can thus infer that at least under fully coherent transport including tunneling, holes will not be able to gain sufficient kinetic energy to achieve impact ionization. 

\begin{figure}[b]
\includegraphics[width=0.45\textwidth]{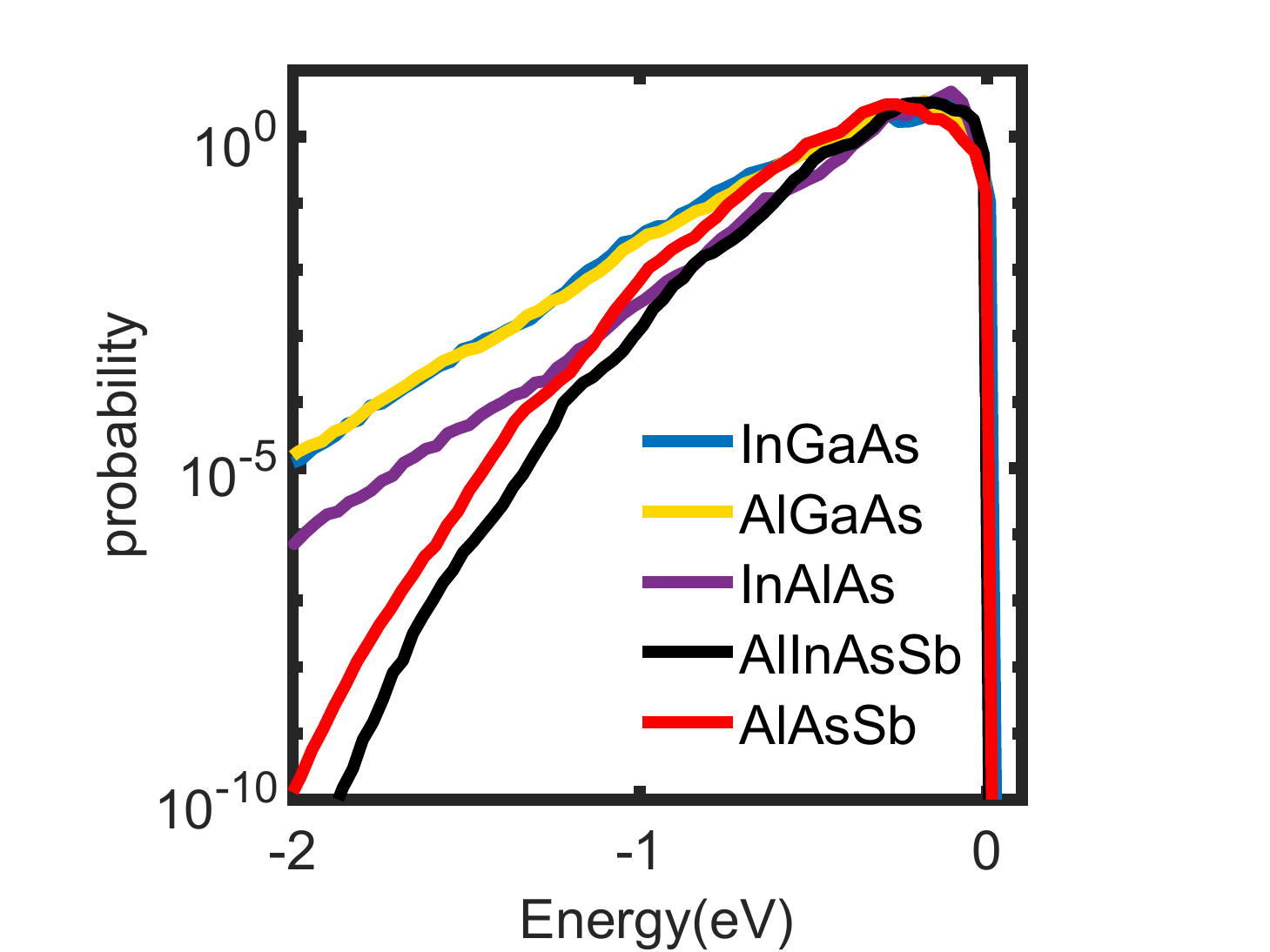}
\caption{\label{fig:bte} Carrier Occupation Probability vs. Energy for the valence band in the presence of optical phonon scattering computed using BTE simulation. InAlAs, AlInAsSb and AlAsSb have lower occupation probability compared to InGaAs and AlGaAs. This prevents holes from gaining the ionization threshold energy.}
\end{figure}

\begin{table}[b]
\begin{center}
\begin{tabular}{ |c| c c|  }  
 \hline
 \textbf{Material} & $\mu _h$ ($ cm^{2}/Vs $) & $E_{opt}$ ($meV$)\\
 \hline
 InAs & 500 & 30 \\
 \hline
 AlAs & 200 & 50 \\ 
 \hline
 GaAs & 400 & 35\\  
 \hline
 AlSb & 400 & 42\\
 \hline
\end{tabular}
\end{center}
\caption{Electron/hole mobilities and optical phonon energies of binary compounds that form the digital alloys \cite{piprek2013semiconductor,shur1996handbook}.}
\label{table:2}
\end{table}

Besides tunneling processes it is possible for carriers to jump across energy gaps through inelastic scattering. In APDs, the dominant scattering mechanism is intervalley optical phonon scattering. Using  the BTE model described in Section \ref{BTE}, the effect of phonon scattering in digital alloys is studied. The carrier mobilities and optical phonon energies of the binary constituents of the alloys used in the BTE simulations are listed in Table ~\ref{table:2}. An effective scattering strength $H_{\vec{p}, \vec{p}'}$ is obtained from the mobility values as described in Section \ref{BTE}. For our BTE simulations, we use the heavy-hole effective masses outlined in Table.~\ref{table:1}. We compute the carrier occupation probability in the valence band under a high electric field of $1 MV/cm$, by solving the three-dimensional Boltzmann equation with the entire set of tight  binding  energy bands within the Brillouin zone of the digital alloy. The optical phonon energy and mobilities  of each alloy are taken to be the average of the binary constituent optical phonon energies and their mobilities. The energy resolved carrier occupation probability for all the alloys is shown in Fig.~\ref{fig:bte}. The valence band plot in Fig.~\ref{fig:bte} shows that the occupation probability for InAlAs, AlInAsSb and AlAsAsb is lower than the other two alloys at high energies. The optical phonon energies of these alloys are not sufficiently large to overcome their minigaps and thus prevent holes from ramping their kinetic energies up to $E_{TH}$. 

\begin{figure}
\includegraphics[width=0.45\textwidth]{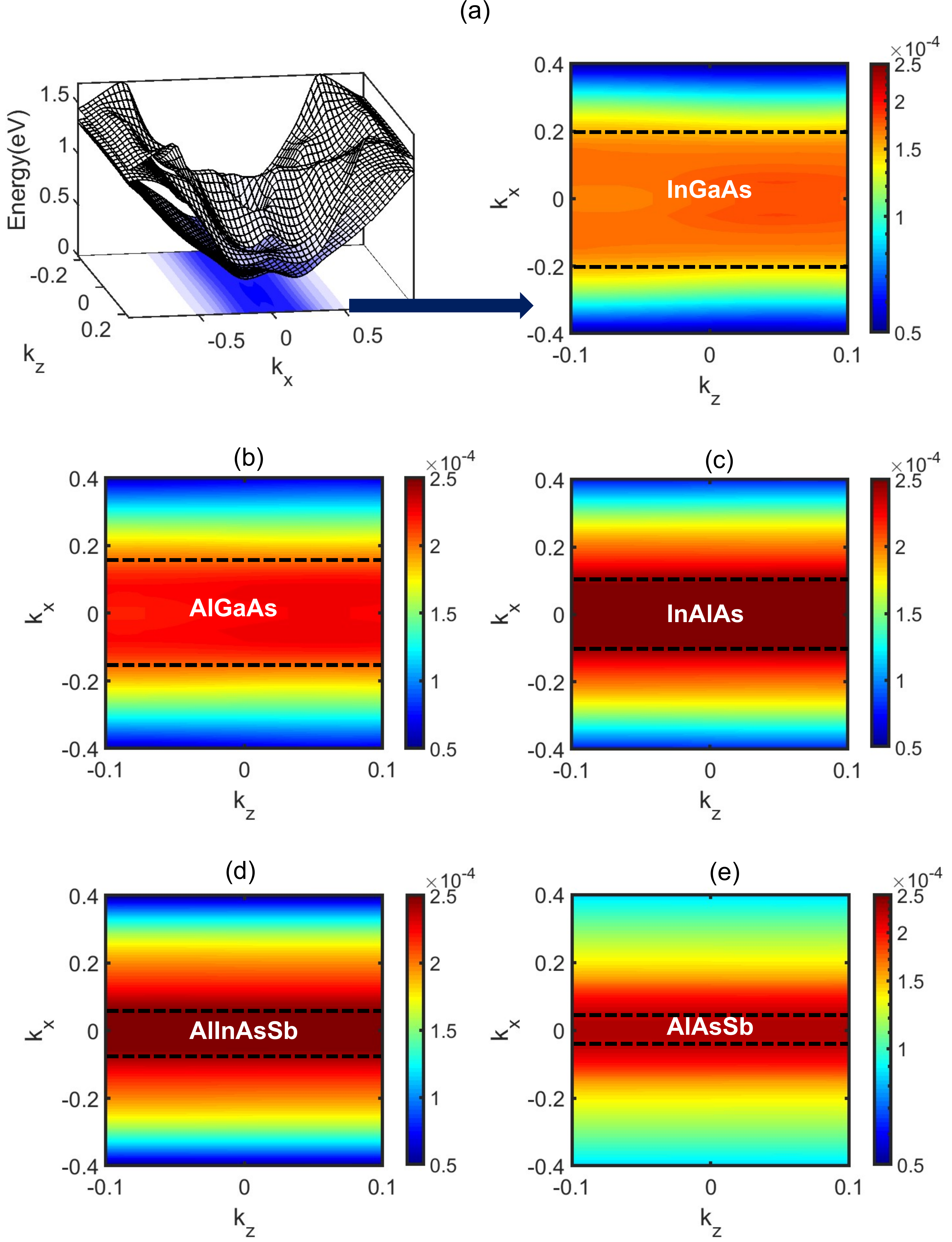}
\caption{\label{fig:3D carrier} Carrier density distribution for (a) InGaAs, (b) AlGaAs, (c) InAlAs, (d) AlInAsSb and (e) AlAsSb.   
}
\end{figure}

The top few valence bands of InGaAs are shown on the left side of Fig.~\ref{fig:3D carrier}(a) and the valence band carrier density distribution is projected onto the bottom. The bands are inverted for better view. For clearer understanding, the InGaAs carrier density distribution contour is also shown on the right. The valence band carrier distributions for the other alloys are shown in Fig.~\ref{fig:3D carrier}(b) AlGaAs, (c) InAlAs, (d) AlInAsSb and (e) AlAsSb. By studying the contours of each material, we observe that the densities for InAlAs, AlInAsSb and AlAsSb are more localized compared to thåt of AlGaAs and InGaAs. This is once again consistent with the lower hole impact ionization of InAlAs, AlInAsSb and AlAsSb.

For InGaAs and AlGaAs, the bandwidths are large enough to allow both holes and electrons to reach $E_{TH}$ easily.  The resulting values of $k$ for these materials are quite high. Correspondingly, these two alloys have higher excess noise. In contrast in InAlAs, AlInAsSb and AlAsSb, it is easy for electrons to reach the threshold energy, but the holes are confined close to the valence band edge. This results in asymmetric ionization coefficients which give a low $k$, leading in turn to low excess noise. 

Armed with these results, we attempt to paint a clearer picture on how the minigaps and band splitting can reduce the excess noise in APDs. Specifically, we  propose a set of empirical inequalities that can used to judge the excess noise performance of a digital alloy.

\section{Empirical Inequalities}
Based on our experimental results and theoretical calculations, five inequalities are proposed that use only material parameters like effective mass and minigap size obtained from our material bandstructures as inputs. In this paper, the transport is in the $[001]$ direction. Since the minigaps considered lie in the LH band, we use the unfolded LH effective mass value in the $\Gamma-[001]$ direction for the inequalities. The masses are obtained using the relationship $\hbar^2 k^2/2m^*=E(1+\alpha E)$ where $\alpha=[(1-m^*/m_0)^2]/E_G$ \cite{FAWCETT19701963}. In reality, the effective masses are complicated tensors that cannot be included in these empirical inequalities but are captured by the NEGF simulations described in Section.~\ref{NEGF}. A digital alloy material should favor low noise if it satisfies the majority of these inequalities. The four main inequalities are:

\begin{equation}
    {\rm{Inequality~(1)~~~~~}}\Delta E_b/E_{TH}<<1 \nonumber
\end{equation}

\begin{equation}
    {\rm{Inequality~(2)~~~~~}}E_{opt}/\Delta E_m<<1 \nonumber
\end{equation}

\begin{equation}
    {\rm{Inequality~(3)~~~~~}}exp\left(-\frac{4\sqrt{2m_l}\Delta E_m^{3/2}}{3q\hbar F} \right)<<1 \nonumber
\end{equation}

\begin{equation}
    {\rm{Inequality.~(4)~~~~~}}exp\left(-\frac{4\sqrt{2m_l}\Delta E_{LS}^{3/2}}{3q\hbar F} \right)<<1 \nonumber
\end{equation}

\begin{figure}[b]
\includegraphics[width=0.49\textwidth]{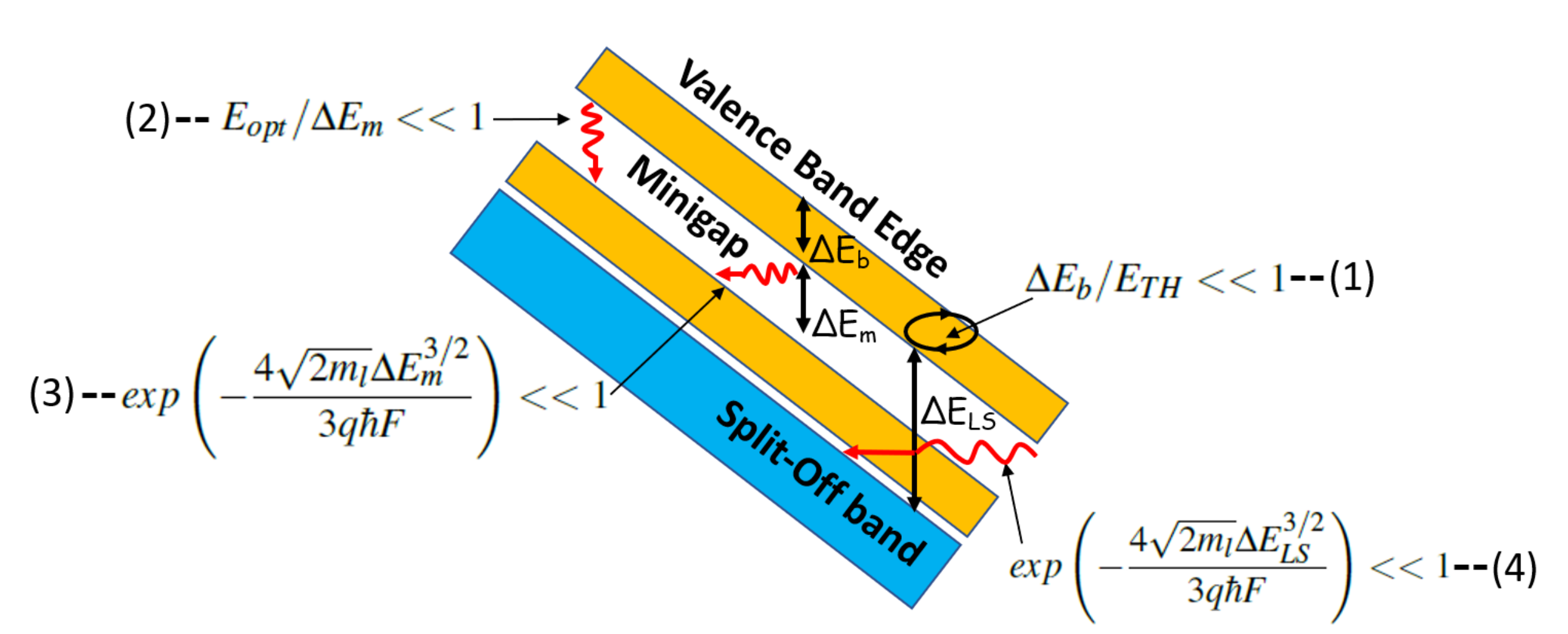}
\caption{\label{fig:inqualities}
Criteria for designing low noise digital alloy APDs. Inequality~(1) states that the bandwidth to the first minigap is lower than the ionization threshold energy. Inequality~(2) asserts that the optical phonon energy has to be less than the minigap size. The tunneling probability for holes to jump across the minigap or from the light-hole band to the split-off band must be low. These are described by Inequality~(3) and Inequality~(4).}
\end{figure}

Here, $\Delta E_b$ represents the energy difference between the VB maximum and the first minigap edge in the VB, $E_{opt}$ is the optical phonon energy and $\Delta E_m$ gives the size of the minigap. The longitudinal effective mass of the band in which the minigap exists is represented by $m_l$. $\Delta E_{LS}$ signifies the energy difference between the LH and SO bands at the $\Gamma$ point. A pictorial view of the different energy differences and inequalities mentioned above is shown in Fig. \ref{fig:inqualities}.  

The  first inequality, Inequality~(1), states that the energy bandwidth $\Delta E_b$ must be less than the ionization threshold energy $E_{TH}$. This means a carrier cannot gain sufficient kinetic energy to impact ionize before reaching the minigap. When a carrier reaches a minigap it faces a barrier (Fig.~\ref{fig:inqualities}), which it can overcome by phonon scattering or  quantum tunneling. Inequality~(2) sets the condition for phonon scattering across the minigap. If the $E_{opt}$ of the material is less than $\Delta E_m$, then the phonon scattering of the carriers across the minigap is inhibited because carriers cannot gain sufficient energy to jump across the gap. It is possible for the carrier to still overcome the minigap by tunneling, and the condition for that is given in Inequality~(3), in terms of the tunneling probability across the minigap under the influence of an electric field. To compute the tunneling probability we consider a triangular barrier in the minigap region and use the well-known Fowler-Nordheim equation. Together Inequalities~(2) and (3) give the effectiveness of the minigap in limiting hole ionization in digital alloys.

\begin{figure}[t]
\includegraphics[width=0.45\textwidth]{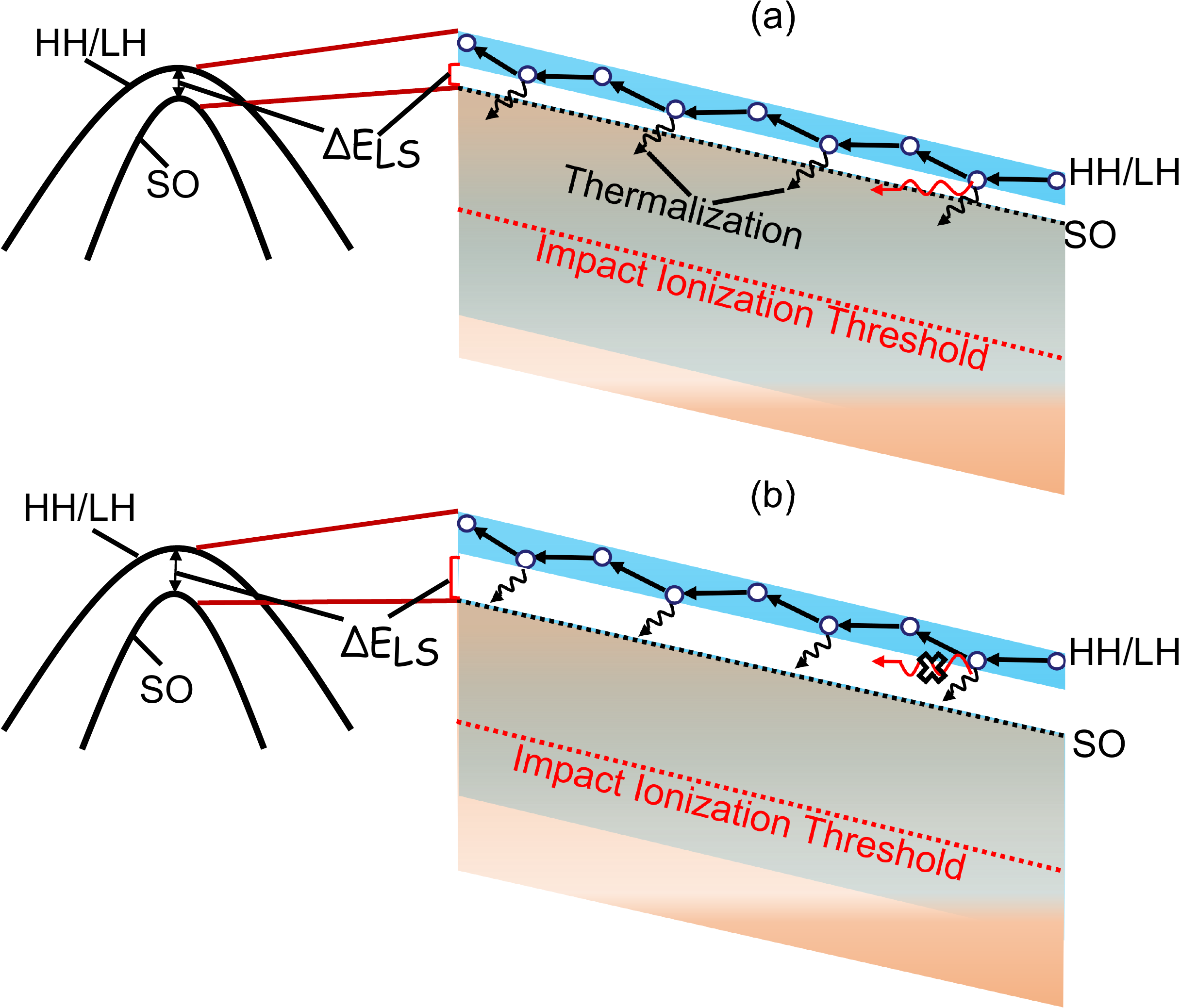}
\caption{\label{fig:LH_SO inequality}
Effect of spin-orbit coupling on LH/SO separation. (a) Weak coupling results in small $\Delta E_{LS}$ and (b) strong coupling results in large $\Delta E_{LS}$.}
\end{figure}

Electron injected digital alloys can in fact achieve low noise even in the absence of minigaps, for instance in a material with a large separation $\Delta E_{LS}$ between the LH and SO bands, like AlAsSb. Holes within HH/LH bands are limited near the valence band edge by thermalization (hole-phonon scattering) due to the heavy effective masses in these bands, preventing them from reaching the ionization threshold energy within the band. An alternate pathway to ionization involves the split-off band. Since the split-off band has a low effective mass, holes require much smaller momentum to reach higher energies in this band, so that holes entering this band from HH/LH can quickly gain their ionization threshold energy. The separation between HH/LH and SO bands is controlled by spin-orbit coupling, as shown in Fig.~\ref{fig:LH_SO inequality}. Strong spin-orbit coupling due to inclusion of heavy elements, like antimony or bismuth, can increase the separation $\Delta E_{LS}$, as shown in Fig.~\ref{fig:LH_SO inequality}(b).  When $\Delta E_{LS}$ is large it becomes very  difficult for holes to reach the threshold energy. Inequality~(4) is accordingly important for APDs in which electron impact ionization is the dominant process, and is a measure of hole tunneling from the light-hole to the split-off band. 

An inherent fifth inequality, satisfied by these five alloys, is 
\begin{equation}
   E_{SC}<E_{TH} 
\label{ineq5}
\end{equation} 

$E_{SC}$ is the energy gained by a hole between successive phonon scattering events, expressed as $E_{SC}=\lambda_{mfp}/F$. The \textit{z}-directed mean free path, $\lambda_{mfp}=v_{sat}\tau_{SC}/2$, where $v_{sat}$ is the saturation velocity and $\tau_{SC}$ is the  scattering lifetime. $E_{SC}$ values of the five alloys at electric fields of $100kV/cm$ and $500kV/cm$ are given in Table.~\ref{table:3}. We extract $\tau_{SC}$ for an alloy by assuming a virtual crystal approximation of the component binary alloy scattering times. $\tau_{SC}$ values for InAs, GaAs, AlAs and AlSb are $0.08ps$, $0.09$, $0.08ps$ and $0.11ps$, respectively \cite{yang2020materials}. A similar average is done for the ternary alloy saturation velocities. Due to unavailability of AlSb $v_{sat}$, InAs $v_{sat}$ is used for AlInAsSb and AlAs $v_{sat}$ for AlAsSb.  InAs, GaAs and AlAs $v_{sat}$ values used are $5\times10^4 m/s$, $9\times10^4 m/s$ and $8\times10^4 m/s$, respectively \cite{Palankovski}. 

\begin{table}[b]
\begin{center}
\begin{tabular}{|m{1.5cm}|m{2cm}|m{2cm}|m{1.5cm}|}
 \hline
Material & $E_{SC}$ (eV) & $E_{SC}$ (eV) & $E_{TH}$ (eV)\\
& at $100kV/cm$ & at $500kV/cm$ &\\
\hline
InGaAs & 0.029 & 0.149 &  0.95 \\ 
 \hline
AlGaAs & 0.036 & 0.181 & 3.91 \\
\hline
InAlAs & 0.028 &  0.138 & 1.85 \\
\hline
AlInAsSb & 0.024 & 0.119 & 1.79 \\
\hline 
AlAsSb & 0.038 & 0.19 & 2.4 \\
\hline
\end{tabular}
\end{center}
\caption{$E_{SC}$ values at $F=100kV/cm$ and $F=500kV/cm$, and $E_{TH}$ of the five alloys. For a material with equal conduction and valence band effective masses, considering parabolic bands, the threshold energy $E_{TH}=1.5 E_G$ \cite{ridley2013quantum}. The same assumption is made here for the fifth inequality as this standard practice in the APD literature.}
\label{table:3}
\end{table}

\begin{table*}[t]
\begin{center}
\begin{tabular}{|c|c|c|c|c|c|c|}  
\hline
 \textbf{Material} & \textbf{Inequality 1} & \textbf{Inequality 2} & \textbf{Inequality 3} & \textbf{Inequality 4} & \textbf{$k$ (digital alloy)} & \textbf{$k$ (random alloy)} \\
\hline
 InGaAs & \cellcolor{teagreen} 0.38 & \cellcolor{cinnabar} 1.08 & \cellcolor{cinnabar} 0.88  &  \cellcolor{cinnabar} 0.006 & \cellcolor{cinnabar} 0.3\cite{InGaAs_expt}  & \cellcolor{cinnabar} 0.5\cite{InGaAs_expt} \\
 \hline
AlGaAs & \cellcolor{bittersweet} 1 & \cellcolor{bittersweet} $\infty$ & \cellcolor{bittersweet} 1 & \cellcolor{bittersweet} $7.2 \times 10^{-4}$ & \cellcolor{bittersweet} 0.1\cite{AlGaAs_expt} & \cellcolor{bittersweet} 0.2\cite{AlGaAs_expt} \\  
 \hline
  InAlAs & \cellcolor{ufogreen} 0.16 & \cellcolor{ufogreen} 0.33 & \cellcolor{ufogreen} 0.17 & \cellcolor{teagreen} $5.6 \times 10^{-4}$  & \cellcolor{teagreen} 0.05\cite{InAlAs_expt} & \cellcolor{bittersweet} 0.2\cite{InAlAs_expt} \\  
 \hline
  AlInAsSb & \cellcolor{ufogreen} 0.17  & \cellcolor{teagreen} 0.59   & \cellcolor{teagreen} 0.53  & \cellcolor{turquoisegreen} $7.9 \times 10^{-7}$ & \cellcolor{turquoisegreen} 0.01\cite{AlInAsSb_expt} & \cellcolor{ufogreen} 0.018\cite{AlInAsSb_expt2} \\
 \hline
  AlAsSb & \cellcolor{turquoisegreen}  0.22 & \cellcolor{turquoisegreen} 0.45 & \cellcolor{turquoisegreen} 0.3 & \cellcolor{ufogreen} $3.4 \times 10^{-7}$ & \cellcolor{ufogreen} 0.005\cite{AlAsSb_expt} & \cellcolor{turquoisegreen} 0.05\cite{AlAsSb_expt2} \\  
  \hline
\end{tabular}
\end{center}
\caption{Suitability of digital alloys for attaining low noise is judged using the proposed inequalities. Here, the color green means beneficial for low noise and red indicates it is detrimental. The impact of the inequality in determining the experimentally determined ionization coefficient ratio $k$ of the material is depicted by the color shades. A darker shade indicates that the inequality has a greater impact on the value of $k$. The experimental random alloy $k$'s of the five alloys are given in column 6.}
\label{table:4}
\end{table*} 

In order to validate these inequalities as design criteria, we apply them to the set of digital alloys mentioned in this paper. We consider a high electric field of $1MV/cm$ for Inequalities~(3) and (4). The values of the left sides of the inequalities for the five alloys- InGaAs, AlGaAs, InAlAs, AlInAsSb and AlAsSb, are given in the first four columns of Table \ref{table:3}, while the measured $k$ is provided as reference in column 5. The table cells are colored green or red. Green cells aid in noise suppression (left sides of the inequalities are relatively small) and red is detrimental to reducing noise (left sides larger and corresponding inequalities not satisfied). Additionally, the color intensities highlight the strength of that inequality (how far the left side is from equality with the right side). A lighter shade represents a smaller impact while a darker shade means that condition has a greater effect on the impact ionization noise. For example, in the case of InGaAs, Inequality~(1) is shaded light green which means it does not effect noise performance significantly. However, the remaining inequalities for InGaAs are shaded dark red, indicating their key role in the high noise and hence high $k$ of InGaAs. The inequalities for AlGaAs, which has a slightly lower $k$, have a lighter shade of red. There are no minigaps for AlGaAs in the light-hole band. There is a minigap in the SO band of AlGaAs which is very deep in the valence band and there are other available states at that energy. Thus, holes can gain sufficient momentum to jump to other bands and bypass the minigap. So, we consider $\Delta E_m=0$ for it. We accordingly expect that AlGaAs has a lower noise. However, since the LH effective mass for AlGaAs is greater than InGaAs, it has lower hole impact ionization and thus lower noise compared to InGaAs. The remaining alloys have significantly lower noise compared to these two. 

The boxes for InAlAs, AlInAsSb and AlAsSb are all green. This means these three alloys are quite favorable for attaining low excess performance. InAlAs has a minigap size $\Delta E_m=0.12 eV$ which is larger than its optical phonon energy. It also has a large LH effective mass which prevents quantum tunneling across the minigap, as well as the LH-SO separation $\Delta E_{LS}$ which is comparable to that of AlGaAs and InGaAs. AlInAsSb has a low value for Inqequality~(1), so that box is shaded dark green. However, for Inequalities~(2) and (3) the values for AlInAsSb are higher than that of InAlAs and are thus shaded in a lighter color. AlInAsSb has a larger LH-SO separation than InAlAs and hence its Inequality~(4) has a darker shade.  In AlAsSb, the values for Inequalities~(1)-(3) have medium shades as they lie between the maximum and minimum values in each of these columns for the corresponding inequalities. However, AlAsSb has a large $\Delta E_{LS}=0.54eV$, so its Inequality~(4) is shaded dark green. Based on the inequality values it would seem InAlAs would have the lowest noise since it has the  darkest shades. However, looking at the Inequality~(4) values for these three materials we can infer that the LH-SO separation plays a critical role in reducing noise. Here, AlAsSb has the lowest $k=0.005$ and also the largest $\Delta E_{LS}$. On the contrary, InAlAs has the highest $k=0.1$ and the smallest $\Delta E_{LS}$. Finally, inequality 5, discussed in the context of split-off states (Eq.~\ref{ineq5}), is trivially satisfied by all five studied alloys. While important, it is thus not tabulated here, as it does not alter the status quo.

In short, the values of the inequalities in Table~\ref{table:3} give a fairly good understanding of the excess noise performance of the set of digital alloys considered in this paper. They can potentially serve as empirical design criteria for judging new digital alloys in consideration as potential material candidates for digital alloy superlattice APDs.

\section{Conclusion}
In this paper, we have studied the digital alloy valence band carrier transport using NEGF and BTE formalisms. Based on our simulation results, we explain how minigaps and LH/SO offset impede hole impact ionization in APDs and improve their excess noise performance.  When these gaps/offsets are sufficiently large they cannot bridged across by quantum tunneling or phonon scattering processes. Furthermore, we propose five inequalities as empirical design criteria for digital alloys with low noise performance capabilities. Material parameters calculated computationally are used as inputs for these. We validate these criteria by explaining the excess noise performance of several experimentally fabricated digital alloy APDs. The design criteria can be used to computationally design new digital alloy structures and benchmark them before actually fabricating these.

\section*{Acknowledgment}
This work was funded by National Science Foundation grant NSF 1936016. The authors thank Dr. John P David of University of Sheffield and Dr. Seth R. Bank of University of Texas-Austin for important discussions and insights. The calculations are done
using the computational resources from High-Performance
Computing systems at the University of Virginia (Rivanna)
and the Extreme Science and Engineering Discovery Environment (XSEDE), which is supported by National Science Foundation grant number ACI-1548562.

\providecommand{\noopsort}[1]{}\providecommand{\singleletter}[1]{#1}%

\begin{thebibliography}{57}%
\makeatletter
\providecommand \@ifxundefined [1]{%
 \@ifx{#1\undefined}
}%
\providecommand \@ifnum [1]{%
 \ifnum #1\expandafter \@firstoftwo
 \else \expandafter \@secondoftwo
 \fi
}%
\providecommand \@ifx [1]{%
 \ifx #1\expandafter \@firstoftwo
 \else \expandafter \@secondoftwo
 \fi
}%
\providecommand \natexlab [1]{#1}%
\providecommand \enquote  [1]{``#1''}%
\providecommand \bibnamefont  [1]{#1}%
\providecommand \bibfnamefont [1]{#1}%
\providecommand \citenamefont [1]{#1}%
\providecommand \href@noop [0]{\@secondoftwo}%
\providecommand \href [0]{\begingroup \@sanitize@url \@href}%
\providecommand \@href[1]{\@@startlink{#1}\@@href}%
\providecommand \@@href[1]{\endgroup#1\@@endlink}%
\providecommand \@sanitize@url [0]{\catcode `\\12\catcode `\$12\catcode
  `\&12\catcode `\#12\catcode `\^12\catcode `\_12\catcode `\%12\relax}%
\providecommand \@@startlink[1]{}%
\providecommand \@@endlink[0]{}%
\providecommand \url  [0]{\begingroup\@sanitize@url \@url }%
\providecommand \@url [1]{\endgroup\@href {#1}{\urlprefix }}%
\providecommand \urlprefix  [0]{URL }%
\providecommand \Eprint [0]{\href }%
\providecommand \doibase [0]{http://dx.doi.org/}%
\providecommand \selectlanguage [0]{\@gobble}%
\providecommand \bibinfo  [0]{\@secondoftwo}%
\providecommand \bibfield  [0]{\@secondoftwo}%
\providecommand \translation [1]{[#1]}%
\providecommand \BibitemOpen [0]{}%
\providecommand \bibitemStop [0]{}%
\providecommand \bibitemNoStop [0]{.\EOS\space}%
\providecommand \EOS [0]{\spacefactor3000\relax}%
\providecommand \BibitemShut  [1]{\csname bibitem#1\endcsname}%
\let\auto@bib@innerbib\@empty
\bibitem [{\citenamefont {{Tosi}}\ \emph {et~al.}(2014)\citenamefont {{Tosi}},
  \citenamefont {{Calandri}}, \citenamefont {{Sanzaro}},\ and\ \citenamefont
  {{Acerbi}}}]{Tosi}%
  \BibitemOpen
  \bibfield  {author} {\bibinfo {author} {\bibfnamefont {A.}~\bibnamefont
  {{Tosi}}}, \bibinfo {author} {\bibfnamefont {N.}~\bibnamefont {{Calandri}}},
  \bibinfo {author} {\bibfnamefont {M.}~\bibnamefont {{Sanzaro}}}, \ and\
  \bibinfo {author} {\bibfnamefont {F.}~\bibnamefont {{Acerbi}}},\ }\href
  {\doibase 10.1109/JSTQE.2014.2328440} {\bibfield  {journal} {\bibinfo
  {journal} {IEEE Journal of Selected Topics in Quantum Electronics}\ }\textbf
  {\bibinfo {volume} {20}},\ \bibinfo {pages} {192} (\bibinfo {year}
  {2014})}\BibitemShut {NoStop}%
\bibitem [{\citenamefont {Campbell}(2008)}]{CAMPBELL2008221}%
  \BibitemOpen
  \bibfield  {author} {\bibinfo {author} {\bibfnamefont {J.~C.}\ \bibnamefont
  {Campbell}},\ }\enquote {\bibinfo {title} {8 - advances in photodetectors},}\
  in\ \href {\doibase https://doi.org/10.1016/B978-0-12-374171-4.00008-3}
  {\emph {\bibinfo {booktitle} {Optical Fiber Telecommunications V A (Fifth
  Edition)}}},\ \bibinfo {series and number} {Optics and Photonics},\ \bibinfo
  {editor} {edited by\ \bibinfo {editor} {\bibfnamefont {I.~P.}\ \bibnamefont
  {Kaminow}}, \bibinfo {editor} {\bibfnamefont {T.}~\bibnamefont {Li}}, \ and\
  \bibinfo {editor} {\bibfnamefont {A.~E.}\ \bibnamefont {Willner}}}\ (\bibinfo
   {publisher} {Academic Press},\ \bibinfo {address} {Burlington},\ \bibinfo
  {year} {2008})\ pp.\ \bibinfo {pages} {221 -- 268},\ \bibinfo {edition}
  {fifth edition}\ ed.\BibitemShut {Stop}%
\bibitem [{\citenamefont {Bertone}\ and\ \citenamefont
  {Clark}(2007)}]{bertone2007avalanche}%
  \BibitemOpen
  \bibfield  {author} {\bibinfo {author} {\bibfnamefont {N.}~\bibnamefont
  {Bertone}}\ and\ \bibinfo {author} {\bibfnamefont {W.}~\bibnamefont
  {Clark}},\ }\href@noop {} {\bibfield  {journal} {\bibinfo  {journal} {Laser
  Focus World}\ }\textbf {\bibinfo {volume} {43}} (\bibinfo {year}
  {2007})}\BibitemShut {NoStop}%
\bibitem [{\citenamefont {Mitra}\ \emph {et~al.}(2006)\citenamefont {Mitra},
  \citenamefont {Beck}, \citenamefont {Skokan}, \citenamefont {Robinson},
  \citenamefont {Antoszewski}, \citenamefont {Winchester}, \citenamefont
  {Keating}, \citenamefont {Nguyen}, \citenamefont {Silva}, \citenamefont
  {Musca}, \citenamefont {Dell},\ and\ \citenamefont {Faraone}}]{Mitra2006}%
  \BibitemOpen
  \bibfield  {author} {\bibinfo {author} {\bibfnamefont {P.}~\bibnamefont
  {Mitra}}, \bibinfo {author} {\bibfnamefont {J.~D.}\ \bibnamefont {Beck}},
  \bibinfo {author} {\bibfnamefont {M.~R.}\ \bibnamefont {Skokan}}, \bibinfo
  {author} {\bibfnamefont {J.~E.}\ \bibnamefont {Robinson}}, \bibinfo {author}
  {\bibfnamefont {J.}~\bibnamefont {Antoszewski}}, \bibinfo {author}
  {\bibfnamefont {K.~J.}\ \bibnamefont {Winchester}}, \bibinfo {author}
  {\bibfnamefont {A.~J.}\ \bibnamefont {Keating}}, \bibinfo {author}
  {\bibfnamefont {T.}~\bibnamefont {Nguyen}}, \bibinfo {author} {\bibfnamefont
  {K.~K. M. B.~D.}\ \bibnamefont {Silva}}, \bibinfo {author} {\bibfnamefont
  {C.~A.}\ \bibnamefont {Musca}}, \bibinfo {author} {\bibfnamefont {J.~M.}\
  \bibnamefont {Dell}}, \ and\ \bibinfo {author} {\bibfnamefont
  {L.}~\bibnamefont {Faraone}},\ }in\ \href {\doibase 10.1117/12.673010} {\emph
  {\bibinfo {booktitle} {Intelligent Integrated Microsystems}}},\ Vol.\
  \bibinfo {volume} {6232},\ \bibinfo {editor} {edited by\ \bibinfo {editor}
  {\bibfnamefont {R.~A.}\ \bibnamefont {Athale}}\ and\ \bibinfo {editor}
  {\bibfnamefont {J.~C.}\ \bibnamefont {Zolper}}},\ \bibinfo {organization}
  {International Society for Optics and Photonics}\ (\bibinfo  {publisher}
  {SPIE},\ \bibinfo {year} {2006})\ pp.\ \bibinfo {pages} {70 --
  80}\BibitemShut {NoStop}%
\bibitem [{\citenamefont {Williams}(2017)}]{Williams2017}%
  \BibitemOpen
  \bibfield  {author} {\bibinfo {author} {\bibfnamefont {G.~M.}\ \bibnamefont
  {Williams}},\ }\href {\doibase 10.1117/1.OE.56.3.031224} {\bibfield
  {journal} {\bibinfo  {journal} {Optical Engineering}\ }\textbf {\bibinfo
  {volume} {56}},\ \bibinfo {pages} {1 } (\bibinfo {year} {2017})}\BibitemShut
  {NoStop}%
\bibitem [{\citenamefont {Nada}\ \emph {et~al.}(2020)\citenamefont {Nada},
  \citenamefont {Nakajima}, \citenamefont {Yoshimatsu}, \citenamefont
  {Nakanishi}, \citenamefont {Tatsumi}, \citenamefont {Yamada}, \citenamefont
  {Sano},\ and\ \citenamefont {Matsuzaki}}]{Nada2020}%
  \BibitemOpen
  \bibfield  {author} {\bibinfo {author} {\bibfnamefont {M.}~\bibnamefont
  {Nada}}, \bibinfo {author} {\bibfnamefont {F.}~\bibnamefont {Nakajima}},
  \bibinfo {author} {\bibfnamefont {T.}~\bibnamefont {Yoshimatsu}}, \bibinfo
  {author} {\bibfnamefont {Y.}~\bibnamefont {Nakanishi}}, \bibinfo {author}
  {\bibfnamefont {S.}~\bibnamefont {Tatsumi}}, \bibinfo {author} {\bibfnamefont
  {Y.}~\bibnamefont {Yamada}}, \bibinfo {author} {\bibfnamefont
  {K.}~\bibnamefont {Sano}}, \ and\ \bibinfo {author} {\bibfnamefont
  {H.}~\bibnamefont {Matsuzaki}},\ }\href {\doibase 10.1063/5.0003573}
  {\bibfield  {journal} {\bibinfo  {journal} {Applied Physics Letters}\
  }\textbf {\bibinfo {volume} {116}},\ \bibinfo {pages} {140502} (\bibinfo
  {year} {2020})},\ \Eprint
  {http://arxiv.org/abs/https://doi.org/10.1063/5.0003573}
  {https://doi.org/10.1063/5.0003573} \BibitemShut {NoStop}%
\bibitem [{\citenamefont {{Pasquinelli}}\ \emph {et~al.}(2020)\citenamefont
  {{Pasquinelli}}, \citenamefont {{Lussana}}, \citenamefont {{Tisa}},
  \citenamefont {{Villa}},\ and\ \citenamefont {{Zappa}}}]{Pasquinelli2020}%
  \BibitemOpen
  \bibfield  {author} {\bibinfo {author} {\bibfnamefont {K.}~\bibnamefont
  {{Pasquinelli}}}, \bibinfo {author} {\bibfnamefont {R.}~\bibnamefont
  {{Lussana}}}, \bibinfo {author} {\bibfnamefont {S.}~\bibnamefont {{Tisa}}},
  \bibinfo {author} {\bibfnamefont {F.}~\bibnamefont {{Villa}}}, \ and\
  \bibinfo {author} {\bibfnamefont {F.}~\bibnamefont {{Zappa}}},\ }\href
  {\doibase 10.1109/JSEN.2020.2977775} {\bibfield  {journal} {\bibinfo
  {journal} {IEEE Sensors Journal}\ }\textbf {\bibinfo {volume} {20}},\
  \bibinfo {pages} {7021} (\bibinfo {year} {2020})}\BibitemShut {NoStop}%
\bibitem [{\citenamefont {Li}\ \emph {et~al.}(2018)\citenamefont {Li},
  \citenamefont {Da~Xu},\ and\ \citenamefont {Zhao}}]{li20185g}%
  \BibitemOpen
  \bibfield  {author} {\bibinfo {author} {\bibfnamefont {S.}~\bibnamefont
  {Li}}, \bibinfo {author} {\bibfnamefont {L.}~\bibnamefont {Da~Xu}}, \ and\
  \bibinfo {author} {\bibfnamefont {S.}~\bibnamefont {Zhao}},\ }\href@noop {}
  {\bibfield  {journal} {\bibinfo  {journal} {Journal of Industrial Information
  Integration}\ }\textbf {\bibinfo {volume} {10}},\ \bibinfo {pages} {1}
  (\bibinfo {year} {2018})}\BibitemShut {NoStop}%
\bibitem [{\citenamefont {Chowdhury}\ \emph {et~al.}(2019)\citenamefont
  {Chowdhury}, \citenamefont {Hasan}, \citenamefont {Shahjalal}, \citenamefont
  {Shin},\ and\ \citenamefont {Jang}}]{Chowdhury5G}%
  \BibitemOpen
  \bibfield  {author} {\bibinfo {author} {\bibfnamefont {M.~Z.}\ \bibnamefont
  {Chowdhury}}, \bibinfo {author} {\bibfnamefont {M.~K.}\ \bibnamefont
  {Hasan}}, \bibinfo {author} {\bibfnamefont {M.}~\bibnamefont {Shahjalal}},
  \bibinfo {author} {\bibfnamefont {E.~B.}\ \bibnamefont {Shin}}, \ and\
  \bibinfo {author} {\bibfnamefont {Y.~M.}\ \bibnamefont {Jang}},\ }in\
  \href@noop {} {\emph {\bibinfo {booktitle} {2019 ICAIIC}}}\ (\bibinfo
  {organization} {IEEE},\ \bibinfo {year} {2019})\ pp.\ \bibinfo {pages}
  {004--007}\BibitemShut {NoStop}%
\bibitem [{\citenamefont {{Liu}}\ and\ \citenamefont
  {{Bowers}}(2018)}]{Liu_III_V}%
  \BibitemOpen
  \bibfield  {author} {\bibinfo {author} {\bibfnamefont {A.~Y.}\ \bibnamefont
  {{Liu}}}\ and\ \bibinfo {author} {\bibfnamefont {J.}~\bibnamefont
  {{Bowers}}},\ }\href {\doibase 10.1109/JSTQE.2018.2854542} {\bibfield
  {journal} {\bibinfo  {journal} {IEEE Journal of Selected Topics in Quantum
  Electronics}\ }\textbf {\bibinfo {volume} {24}},\ \bibinfo {pages} {1}
  (\bibinfo {year} {2018})}\BibitemShut {NoStop}%
\bibitem [{\citenamefont {{Campbell}}(2016)}]{apd_recent}%
  \BibitemOpen
  \bibfield  {author} {\bibinfo {author} {\bibfnamefont {J.~C.}\ \bibnamefont
  {{Campbell}}},\ }\href {\doibase 10.1109/JLT.2015.2453092} {\bibfield
  {journal} {\bibinfo  {journal} {Journal of Lightwave Technology}\ }\textbf
  {\bibinfo {volume} {34}},\ \bibinfo {pages} {278} (\bibinfo {year}
  {2016})}\BibitemShut {NoStop}%
\bibitem [{\citenamefont {McIntyre}(1966)}]{mcintyre1966multiplication}%
  \BibitemOpen
  \bibfield  {author} {\bibinfo {author} {\bibfnamefont {R.}~\bibnamefont
  {McIntyre}},\ }\href@noop {} {\bibfield  {journal} {\bibinfo  {journal} {IEEE
  Transactions on Electron Devices}\ ,\ \bibinfo {pages} {164}} (\bibinfo
  {year} {1966})}\BibitemShut {NoStop}%
\bibitem [{\citenamefont {{Teich}}\ \emph {et~al.}(1986)\citenamefont
  {{Teich}}, \citenamefont {{Matsuo}},\ and\ \citenamefont
  {{Saleh}}}]{Teich1986}%
  \BibitemOpen
  \bibfield  {author} {\bibinfo {author} {\bibfnamefont {M.}~\bibnamefont
  {{Teich}}}, \bibinfo {author} {\bibfnamefont {K.}~\bibnamefont {{Matsuo}}}, \
  and\ \bibinfo {author} {\bibfnamefont {B.}~\bibnamefont {{Saleh}}},\ }\href
  {\doibase 10.1109/JQE.1986.1073137} {\bibfield  {journal} {\bibinfo
  {journal} {IEEE Journal of Quantum Electronics}\ }\textbf {\bibinfo {volume}
  {22}},\ \bibinfo {pages} {1184} (\bibinfo {year} {1986})}\BibitemShut
  {NoStop}%
\bibitem [{\citenamefont {{Hakim}}\ \emph {et~al.}(1990)\citenamefont
  {{Hakim}}, \citenamefont {{Saleh}},\ and\ \citenamefont
  {{Teich}}}]{Teich1990}%
  \BibitemOpen
  \bibfield  {author} {\bibinfo {author} {\bibfnamefont {N.~Z.}\ \bibnamefont
  {{Hakim}}}, \bibinfo {author} {\bibfnamefont {B.~E.~A.}\ \bibnamefont
  {{Saleh}}}, \ and\ \bibinfo {author} {\bibfnamefont {M.~C.}\ \bibnamefont
  {{Teich}}},\ }\href {\doibase 10.1109/16.47763} {\bibfield  {journal}
  {\bibinfo  {journal} {IEEE Transactions on Electron Devices}\ }\textbf
  {\bibinfo {volume} {37}},\ \bibinfo {pages} {599} (\bibinfo {year}
  {1990})}\BibitemShut {NoStop}%
\bibitem [{\citenamefont {Zheng}\ \emph
  {et~al.}(2019{\natexlab{a}})\citenamefont {Zheng}, \citenamefont {Tan},
  \citenamefont {Yuan}, \citenamefont {Ghosh},\ and\ \citenamefont
  {Campbell}}]{InAlAs_expt}%
  \BibitemOpen
  \bibfield  {author} {\bibinfo {author} {\bibfnamefont {J.}~\bibnamefont
  {Zheng}}, \bibinfo {author} {\bibfnamefont {Y.}~\bibnamefont {Tan}}, \bibinfo
  {author} {\bibfnamefont {Y.}~\bibnamefont {Yuan}}, \bibinfo {author}
  {\bibfnamefont {A.}~\bibnamefont {Ghosh}}, \ and\ \bibinfo {author}
  {\bibfnamefont {J.}~\bibnamefont {Campbell}},\ }\href@noop {} {\bibfield
  {journal} {\bibinfo  {journal} {Journal of Applied Physics}\ }\textbf
  {\bibinfo {volume} {125}},\ \bibinfo {pages} {082514} (\bibinfo {year}
  {2019}{\natexlab{a}})}\BibitemShut {NoStop}%
\bibitem [{\citenamefont {Bank}\ \emph {et~al.}(2018)\citenamefont {Bank},
  \citenamefont {Campbell}, \citenamefont {Maddox}, \citenamefont {Rockwell},
  \citenamefont {Woodson}, \citenamefont {Ren}, \citenamefont {Jones},
  \citenamefont {March}, \citenamefont {Zheng},\ and\ \citenamefont
  {Yuan}}]{AlInAsSb_expt}%
  \BibitemOpen
  \bibfield  {author} {\bibinfo {author} {\bibfnamefont {S.~R.}\ \bibnamefont
  {Bank}}, \bibinfo {author} {\bibfnamefont {J.~C.}\ \bibnamefont {Campbell}},
  \bibinfo {author} {\bibfnamefont {S.~J.}\ \bibnamefont {Maddox}}, \bibinfo
  {author} {\bibfnamefont {A.~K.}\ \bibnamefont {Rockwell}}, \bibinfo {author}
  {\bibfnamefont {M.~E.}\ \bibnamefont {Woodson}}, \bibinfo {author}
  {\bibfnamefont {M.}~\bibnamefont {Ren}}, \bibinfo {author} {\bibfnamefont
  {A.}~\bibnamefont {Jones}}, \bibinfo {author} {\bibfnamefont
  {S.}~\bibnamefont {March}}, \bibinfo {author} {\bibfnamefont
  {J.}~\bibnamefont {Zheng}}, \ and\ \bibinfo {author} {\bibfnamefont
  {Y.}~\bibnamefont {Yuan}},\ }in\ \href@noop {} {\emph {\bibinfo {booktitle}
  {2018 IEEE RAPID}}}\ (\bibinfo {organization} {IEEE},\ \bibinfo {year}
  {2018})\ pp.\ \bibinfo {pages} {1--3}\BibitemShut {NoStop}%
\bibitem [{\citenamefont {Yi}\ \emph {et~al.}(2019)\citenamefont {Yi},
  \citenamefont {Xie}, \citenamefont {Liang}, \citenamefont {Lim},
  \citenamefont {Cheong}, \citenamefont {Debnath}, \citenamefont {Huffaker},
  \citenamefont {Tan},\ and\ \citenamefont {David}}]{AlAsSb_expt}%
  \BibitemOpen
  \bibfield  {author} {\bibinfo {author} {\bibfnamefont {X.}~\bibnamefont
  {Yi}}, \bibinfo {author} {\bibfnamefont {S.}~\bibnamefont {Xie}}, \bibinfo
  {author} {\bibfnamefont {B.}~\bibnamefont {Liang}}, \bibinfo {author}
  {\bibfnamefont {L.~W.}\ \bibnamefont {Lim}}, \bibinfo {author} {\bibfnamefont
  {J.~S.}\ \bibnamefont {Cheong}}, \bibinfo {author} {\bibfnamefont {M.~C.}\
  \bibnamefont {Debnath}}, \bibinfo {author} {\bibfnamefont {D.~L.}\
  \bibnamefont {Huffaker}}, \bibinfo {author} {\bibfnamefont {C.~H.}\
  \bibnamefont {Tan}}, \ and\ \bibinfo {author} {\bibfnamefont {J.~P.}\
  \bibnamefont {David}},\ }\href@noop {} {\bibfield  {journal} {\bibinfo
  {journal} {Nature Photonics}\ }\textbf {\bibinfo {volume} {13}},\ \bibinfo
  {pages} {683} (\bibinfo {year} {2019})}\BibitemShut {NoStop}%
\bibitem [{\citenamefont {Rockwell}\ \emph {et~al.}(2018)\citenamefont
  {Rockwell}, \citenamefont {Ren}, \citenamefont {Woodson}, \citenamefont
  {Jones}, \citenamefont {March}, \citenamefont {Tan}, \citenamefont {Yuan},
  \citenamefont {Sun}, \citenamefont {Hool}, \citenamefont {Maddox} \emph
  {et~al.}}]{InGaAs_expt}%
  \BibitemOpen
  \bibfield  {author} {\bibinfo {author} {\bibfnamefont {A.}~\bibnamefont
  {Rockwell}}, \bibinfo {author} {\bibfnamefont {M.}~\bibnamefont {Ren}},
  \bibinfo {author} {\bibfnamefont {M.}~\bibnamefont {Woodson}}, \bibinfo
  {author} {\bibfnamefont {A.}~\bibnamefont {Jones}}, \bibinfo {author}
  {\bibfnamefont {S.}~\bibnamefont {March}}, \bibinfo {author} {\bibfnamefont
  {Y.}~\bibnamefont {Tan}}, \bibinfo {author} {\bibfnamefont {Y.}~\bibnamefont
  {Yuan}}, \bibinfo {author} {\bibfnamefont {Y.}~\bibnamefont {Sun}}, \bibinfo
  {author} {\bibfnamefont {R.}~\bibnamefont {Hool}}, \bibinfo {author}
  {\bibfnamefont {S.}~\bibnamefont {Maddox}},  \emph {et~al.},\ }\href@noop {}
  {\bibfield  {journal} {\bibinfo  {journal} {Applied Physics Letters}\
  }\textbf {\bibinfo {volume} {113}},\ \bibinfo {pages} {102106} (\bibinfo
  {year} {2018})}\BibitemShut {NoStop}%
\bibitem [{\citenamefont {Yuan}\ \emph {et~al.}(2018)\citenamefont {Yuan},
  \citenamefont {Zheng}, \citenamefont {Tan}, \citenamefont {Peng},
  \citenamefont {Rockwell}, \citenamefont {Bank}, \citenamefont {Ghosh},\ and\
  \citenamefont {Campbell}}]{AlGaAs_expt}%
  \BibitemOpen
  \bibfield  {author} {\bibinfo {author} {\bibfnamefont {Y.}~\bibnamefont
  {Yuan}}, \bibinfo {author} {\bibfnamefont {J.}~\bibnamefont {Zheng}},
  \bibinfo {author} {\bibfnamefont {Y.}~\bibnamefont {Tan}}, \bibinfo {author}
  {\bibfnamefont {Y.}~\bibnamefont {Peng}}, \bibinfo {author} {\bibfnamefont
  {A.-K.}\ \bibnamefont {Rockwell}}, \bibinfo {author} {\bibfnamefont {S.~R.}\
  \bibnamefont {Bank}}, \bibinfo {author} {\bibfnamefont {A.}~\bibnamefont
  {Ghosh}}, \ and\ \bibinfo {author} {\bibfnamefont {J.~C.}\ \bibnamefont
  {Campbell}},\ }\href@noop {} {\bibfield  {journal} {\bibinfo  {journal}
  {Photonics Research}\ }\textbf {\bibinfo {volume} {6}},\ \bibinfo {pages}
  {794} (\bibinfo {year} {2018})}\BibitemShut {NoStop}%
\bibitem [{\citenamefont {Zheng}\ \emph {et~al.}(2020)\citenamefont {Zheng},
  \citenamefont {Ahmed}, \citenamefont {Yuan}, \citenamefont {Jones},
  \citenamefont {Tan}, \citenamefont {Rockwell}, \citenamefont {March},
  \citenamefont {Bank}, \citenamefont {Ghosh},\ and\ \citenamefont
  {Campbell}}]{AlInAsSb_MC}%
  \BibitemOpen
  \bibfield  {author} {\bibinfo {author} {\bibfnamefont {J.}~\bibnamefont
  {Zheng}}, \bibinfo {author} {\bibfnamefont {S.~Z.}\ \bibnamefont {Ahmed}},
  \bibinfo {author} {\bibfnamefont {Y.}~\bibnamefont {Yuan}}, \bibinfo {author}
  {\bibfnamefont {A.}~\bibnamefont {Jones}}, \bibinfo {author} {\bibfnamefont
  {Y.}~\bibnamefont {Tan}}, \bibinfo {author} {\bibfnamefont {A.~K.}\
  \bibnamefont {Rockwell}}, \bibinfo {author} {\bibfnamefont {S.~D.}\
  \bibnamefont {March}}, \bibinfo {author} {\bibfnamefont {S.~R.}\ \bibnamefont
  {Bank}}, \bibinfo {author} {\bibfnamefont {A.~W.}\ \bibnamefont {Ghosh}}, \
  and\ \bibinfo {author} {\bibfnamefont {J.~C.}\ \bibnamefont {Campbell}},\
  }\href {\doibase https://doi.org/10.1002/inf2.12112} {\bibfield  {journal}
  {\bibinfo  {journal} {InfoMat}\ }\textbf {\bibinfo {volume} {2}},\ \bibinfo
  {pages} {1236} (\bibinfo {year} {2020})},\ \Eprint
  {http://arxiv.org/abs/https://onlinelibrary.wiley.com/doi/pdf/10.1002/inf2.12112}
  {https://onlinelibrary.wiley.com/doi/pdf/10.1002/inf2.12112} \BibitemShut
  {NoStop}%
\bibitem [{\citenamefont {Zheng}\ \emph
  {et~al.}(2019{\natexlab{b}})\citenamefont {Zheng}, \citenamefont {Yuan},
  \citenamefont {Tan}, \citenamefont {Peng}, \citenamefont {Rockwell},
  \citenamefont {Bank}, \citenamefont {Ghosh},\ and\ \citenamefont
  {Campbell}}]{InAlAs_MC}%
  \BibitemOpen
  \bibfield  {author} {\bibinfo {author} {\bibfnamefont {J.}~\bibnamefont
  {Zheng}}, \bibinfo {author} {\bibfnamefont {Y.}~\bibnamefont {Yuan}},
  \bibinfo {author} {\bibfnamefont {Y.}~\bibnamefont {Tan}}, \bibinfo {author}
  {\bibfnamefont {Y.}~\bibnamefont {Peng}}, \bibinfo {author} {\bibfnamefont
  {A.}~\bibnamefont {Rockwell}}, \bibinfo {author} {\bibfnamefont {S.~R.}\
  \bibnamefont {Bank}}, \bibinfo {author} {\bibfnamefont {A.~W.}\ \bibnamefont
  {Ghosh}}, \ and\ \bibinfo {author} {\bibfnamefont {J.~C.}\ \bibnamefont
  {Campbell}},\ }\href {\doibase 10.1063/1.5114918} {\bibfield  {journal}
  {\bibinfo  {journal} {Applied Physics Letters}\ }\textbf {\bibinfo {volume}
  {115}},\ \bibinfo {pages} {171106} (\bibinfo {year} {2019}{\natexlab{b}})},\
  \Eprint {http://arxiv.org/abs/https://doi.org/10.1063/1.5114918}
  {https://doi.org/10.1063/1.5114918} \BibitemShut {NoStop}%
\bibitem [{\citenamefont {{Zheng}}\ \emph {et~al.}(2018)\citenamefont
  {{Zheng}}, \citenamefont {{Yuan}}, \citenamefont {{Tan}}, \citenamefont
  {{Peng}}, \citenamefont {{Rockwell}}, \citenamefont {{Bank}}, \citenamefont
  {{Ghosh}},\ and\ \citenamefont {{Campbell}}}]{InAlAs_MC2}%
  \BibitemOpen
  \bibfield  {author} {\bibinfo {author} {\bibfnamefont {J.}~\bibnamefont
  {{Zheng}}}, \bibinfo {author} {\bibfnamefont {Y.}~\bibnamefont {{Yuan}}},
  \bibinfo {author} {\bibfnamefont {Y.}~\bibnamefont {{Tan}}}, \bibinfo
  {author} {\bibfnamefont {Y.}~\bibnamefont {{Peng}}}, \bibinfo {author}
  {\bibfnamefont {A.~K.}\ \bibnamefont {{Rockwell}}}, \bibinfo {author}
  {\bibfnamefont {S.~R.}\ \bibnamefont {{Bank}}}, \bibinfo {author}
  {\bibfnamefont {A.~W.}\ \bibnamefont {{Ghosh}}}, \ and\ \bibinfo {author}
  {\bibfnamefont {J.~C.}\ \bibnamefont {{Campbell}}},\ }\href {\doibase
  10.1109/JLT.2018.2844114} {\bibfield  {journal} {\bibinfo  {journal} {Journal
  of Lightwave Technology}\ }\textbf {\bibinfo {volume} {36}},\ \bibinfo
  {pages} {3580} (\bibinfo {year} {2018})}\BibitemShut {NoStop}%
\bibitem [{\citenamefont {Tan}\ \emph {et~al.}(2016{\natexlab{a}})\citenamefont
  {Tan}, \citenamefont {Povolotskyi}, \citenamefont {Kubis}, \citenamefont
  {Boykin},\ and\ \citenamefont {Klimeck}}]{TanETB}%
  \BibitemOpen
  \bibfield  {author} {\bibinfo {author} {\bibfnamefont {Y.}~\bibnamefont
  {Tan}}, \bibinfo {author} {\bibfnamefont {M.}~\bibnamefont {Povolotskyi}},
  \bibinfo {author} {\bibfnamefont {T.}~\bibnamefont {Kubis}}, \bibinfo
  {author} {\bibfnamefont {T.~B.}\ \bibnamefont {Boykin}}, \ and\ \bibinfo
  {author} {\bibfnamefont {G.}~\bibnamefont {Klimeck}},\ }\href {\doibase
  10.1103/PhysRevB.94.045311} {\bibfield  {journal} {\bibinfo  {journal} {Phys.
  Rev. B}\ }\textbf {\bibinfo {volume} {94}},\ \bibinfo {pages} {045311}
  (\bibinfo {year} {2016}{\natexlab{a}})}\BibitemShut {NoStop}%
\bibitem [{\citenamefont {Towns}\ \emph {et~al.}(2014)\citenamefont {Towns},
  \citenamefont {Cockerill}, \citenamefont {Dahan}, \citenamefont {Foster},
  \citenamefont {Gaither}, \citenamefont {Grimshaw}, \citenamefont {Hazlewood},
  \citenamefont {Lathrop}, \citenamefont {Lifka}, \citenamefont {Peterson},
  \citenamefont {Roskies}, \citenamefont {Scott},\ and\ \citenamefont
  {Wilkins-Diehr}}]{xsede}%
  \BibitemOpen
  \bibfield  {author} {\bibinfo {author} {\bibfnamefont {J.}~\bibnamefont
  {Towns}}, \bibinfo {author} {\bibfnamefont {T.}~\bibnamefont {Cockerill}},
  \bibinfo {author} {\bibfnamefont {M.}~\bibnamefont {Dahan}}, \bibinfo
  {author} {\bibfnamefont {I.}~\bibnamefont {Foster}}, \bibinfo {author}
  {\bibfnamefont {K.}~\bibnamefont {Gaither}}, \bibinfo {author} {\bibfnamefont
  {A.}~\bibnamefont {Grimshaw}}, \bibinfo {author} {\bibfnamefont
  {V.}~\bibnamefont {Hazlewood}}, \bibinfo {author} {\bibfnamefont
  {S.}~\bibnamefont {Lathrop}}, \bibinfo {author} {\bibfnamefont
  {D.}~\bibnamefont {Lifka}}, \bibinfo {author} {\bibfnamefont {G.~D.}\
  \bibnamefont {Peterson}}, \bibinfo {author} {\bibfnamefont {R.}~\bibnamefont
  {Roskies}}, \bibinfo {author} {\bibfnamefont {J.}~\bibnamefont {Scott}}, \
  and\ \bibinfo {author} {\bibfnamefont {N.}~\bibnamefont {Wilkins-Diehr}},\
  }\href {\doibase 10.1109/MCSE.2014.80} {\bibfield  {journal} {\bibinfo
  {journal} {Computing in Science \& Engineering}\ }\textbf {\bibinfo {volume}
  {16}},\ \bibinfo {pages} {62} (\bibinfo {year} {2014})}\BibitemShut {NoStop}%
\bibitem [{\citenamefont {Tan}\ \emph {et~al.}(2015)\citenamefont {Tan},
  \citenamefont {Povolotskyi}, \citenamefont {Kubis}, \citenamefont {Boykin},\
  and\ \citenamefont {Klimeck}}]{TanSi}%
  \BibitemOpen
  \bibfield  {author} {\bibinfo {author} {\bibfnamefont {Y.~P.}\ \bibnamefont
  {Tan}}, \bibinfo {author} {\bibfnamefont {M.}~\bibnamefont {Povolotskyi}},
  \bibinfo {author} {\bibfnamefont {T.}~\bibnamefont {Kubis}}, \bibinfo
  {author} {\bibfnamefont {T.~B.}\ \bibnamefont {Boykin}}, \ and\ \bibinfo
  {author} {\bibfnamefont {G.}~\bibnamefont {Klimeck}},\ }\href {\doibase
  10.1103/PhysRevB.92.085301} {\bibfield  {journal} {\bibinfo  {journal} {Phys.
  Rev. B}\ }\textbf {\bibinfo {volume} {92}},\ \bibinfo {pages} {085301}
  (\bibinfo {year} {2015})}\BibitemShut {NoStop}%
\bibitem [{\citenamefont {Kienle}\ \emph
  {et~al.}(2006{\natexlab{a}})\citenamefont {Kienle}, \citenamefont {Cerda},\
  and\ \citenamefont {Ghosh}}]{huckel_cnt}%
  \BibitemOpen
  \bibfield  {author} {\bibinfo {author} {\bibfnamefont {D.}~\bibnamefont
  {Kienle}}, \bibinfo {author} {\bibfnamefont {J.~I.}\ \bibnamefont {Cerda}}, \
  and\ \bibinfo {author} {\bibfnamefont {A.~W.}\ \bibnamefont {Ghosh}},\ }\href
  {\doibase 10.1063/1.2259818} {\bibfield  {journal} {\bibinfo  {journal}
  {Journal of Applied Physics}\ }\textbf {\bibinfo {volume} {100}},\ \bibinfo
  {pages} {043714} (\bibinfo {year} {2006}{\natexlab{a}})},\ \Eprint
  {http://arxiv.org/abs/https://doi.org/10.1063/1.2259818}
  {https://doi.org/10.1063/1.2259818} \BibitemShut {NoStop}%
\bibitem [{\citenamefont {Kienle}\ \emph
  {et~al.}(2006{\natexlab{b}})\citenamefont {Kienle}, \citenamefont {Bevan},
  \citenamefont {Liang}, \citenamefont {Siddiqui}, \citenamefont {Cerda},\ and\
  \citenamefont {Ghosh}}]{huckel_silicon}%
  \BibitemOpen
  \bibfield  {author} {\bibinfo {author} {\bibfnamefont {D.}~\bibnamefont
  {Kienle}}, \bibinfo {author} {\bibfnamefont {K.~H.}\ \bibnamefont {Bevan}},
  \bibinfo {author} {\bibfnamefont {G.-C.}\ \bibnamefont {Liang}}, \bibinfo
  {author} {\bibfnamefont {L.}~\bibnamefont {Siddiqui}}, \bibinfo {author}
  {\bibfnamefont {J.~I.}\ \bibnamefont {Cerda}}, \ and\ \bibinfo {author}
  {\bibfnamefont {A.~W.}\ \bibnamefont {Ghosh}},\ }\href {\doibase
  10.1063/1.2259820} {\bibfield  {journal} {\bibinfo  {journal} {Journal of
  Applied Physics}\ }\textbf {\bibinfo {volume} {100}},\ \bibinfo {pages}
  {043715} (\bibinfo {year} {2006}{\natexlab{b}})},\ \Eprint
  {http://arxiv.org/abs/https://doi.org/10.1063/1.2259820}
  {https://doi.org/10.1063/1.2259820} \BibitemShut {NoStop}%
\bibitem [{\citenamefont {Heyd}\ \emph {et~al.}(2003)\citenamefont {Heyd},
  \citenamefont {Scuseria},\ and\ \citenamefont {Ernzerhof}}]{heyd2003hybrid}%
  \BibitemOpen
  \bibfield  {author} {\bibinfo {author} {\bibfnamefont {J.}~\bibnamefont
  {Heyd}}, \bibinfo {author} {\bibfnamefont {G.~E.}\ \bibnamefont {Scuseria}},
  \ and\ \bibinfo {author} {\bibfnamefont {M.}~\bibnamefont {Ernzerhof}},\
  }\href@noop {} {\bibfield  {journal} {\bibinfo  {journal} {The Journal of
  chemical physics}\ }\textbf {\bibinfo {volume} {118}},\ \bibinfo {pages}
  {8207} (\bibinfo {year} {2003})}\BibitemShut {NoStop}%
\bibitem [{\citenamefont {Ahmed}\ \emph
  {et~al.}(2018{\natexlab{a}})\citenamefont {Ahmed}, \citenamefont {Tan},
  \citenamefont {Truesdell}, \citenamefont {Calhoun},\ and\ \citenamefont
  {Ghosh}}]{AhmedTFET}%
  \BibitemOpen
  \bibfield  {author} {\bibinfo {author} {\bibfnamefont {S.~Z.}\ \bibnamefont
  {Ahmed}}, \bibinfo {author} {\bibfnamefont {Y.}~\bibnamefont {Tan}}, \bibinfo
  {author} {\bibfnamefont {D.~S.}\ \bibnamefont {Truesdell}}, \bibinfo {author}
  {\bibfnamefont {B.~H.}\ \bibnamefont {Calhoun}}, \ and\ \bibinfo {author}
  {\bibfnamefont {A.~W.}\ \bibnamefont {Ghosh}},\ }\href {\doibase
  10.1063/1.5044434} {\bibfield  {journal} {\bibinfo  {journal} {Journal of
  Applied Physics}\ }\textbf {\bibinfo {volume} {124}},\ \bibinfo {pages}
  {154503} (\bibinfo {year} {2018}{\natexlab{a}})},\ \Eprint
  {http://arxiv.org/abs/https://doi.org/10.1063/1.5044434}
  {https://doi.org/10.1063/1.5044434} \BibitemShut {NoStop}%
\bibitem [{\citenamefont {Tan}\ \emph {et~al.}(2016{\natexlab{b}})\citenamefont
  {Tan}, \citenamefont {Chen},\ and\ \citenamefont {Ghosh}}]{tan_unfolding}%
  \BibitemOpen
  \bibfield  {author} {\bibinfo {author} {\bibfnamefont {Y.}~\bibnamefont
  {Tan}}, \bibinfo {author} {\bibfnamefont {F.~W.}\ \bibnamefont {Chen}}, \
  and\ \bibinfo {author} {\bibfnamefont {A.~W.}\ \bibnamefont {Ghosh}},\
  }\href@noop {} {\bibfield  {journal} {\bibinfo  {journal} {Applied Physics
  Letters}\ }\textbf {\bibinfo {volume} {109}},\ \bibinfo {pages} {101601}
  (\bibinfo {year} {2016}{\natexlab{b}})}\BibitemShut {NoStop}%
\bibitem [{\citenamefont {Boykin}\ \emph
  {et~al.}(2007{\natexlab{a}})\citenamefont {Boykin}, \citenamefont {Kharche},\
  and\ \citenamefont {Klimeck}}]{boykin_unfolding1}%
  \BibitemOpen
  \bibfield  {author} {\bibinfo {author} {\bibfnamefont {T.~B.}\ \bibnamefont
  {Boykin}}, \bibinfo {author} {\bibfnamefont {N.}~\bibnamefont {Kharche}}, \
  and\ \bibinfo {author} {\bibfnamefont {G.}~\bibnamefont {Klimeck}},\
  }\href@noop {} {\bibfield  {journal} {\bibinfo  {journal} {Physical Review
  B}\ }\textbf {\bibinfo {volume} {76}},\ \bibinfo {pages} {035310} (\bibinfo
  {year} {2007}{\natexlab{a}})}\BibitemShut {NoStop}%
\bibitem [{\citenamefont {Boykin}\ \emph
  {et~al.}(2007{\natexlab{b}})\citenamefont {Boykin}, \citenamefont {Kharche},
  \citenamefont {Klimeck},\ and\ \citenamefont
  {Korkusinski}}]{boykin_unfolding2}%
  \BibitemOpen
  \bibfield  {author} {\bibinfo {author} {\bibfnamefont {T.~B.}\ \bibnamefont
  {Boykin}}, \bibinfo {author} {\bibfnamefont {N.}~\bibnamefont {Kharche}},
  \bibinfo {author} {\bibfnamefont {G.}~\bibnamefont {Klimeck}}, \ and\
  \bibinfo {author} {\bibfnamefont {M.}~\bibnamefont {Korkusinski}},\
  }\href@noop {} {\bibfield  {journal} {\bibinfo  {journal} {Journal of
  Physics: Condensed Matter}\ }\textbf {\bibinfo {volume} {19}},\ \bibinfo
  {pages} {036203} (\bibinfo {year} {2007}{\natexlab{b}})}\BibitemShut
  {NoStop}%
\bibitem [{\citenamefont {St{\o}vneng}\ and\ \citenamefont
  {Lipavsk\'y}(1994)}]{stovneng1993multiband}%
  \BibitemOpen
  \bibfield  {author} {\bibinfo {author} {\bibfnamefont {J.~A.}\ \bibnamefont
  {St{\o}vneng}}\ and\ \bibinfo {author} {\bibfnamefont {P.}~\bibnamefont
  {Lipavsk\'y}},\ }\href {\doibase 10.1103/PhysRevB.49.16494} {\bibfield
  {journal} {\bibinfo  {journal} {Phys. Rev. B}\ }\textbf {\bibinfo {volume}
  {49}},\ \bibinfo {pages} {16494} (\bibinfo {year} {1994})}\BibitemShut
  {NoStop}%
\bibitem [{\citenamefont {Stickler}(2013)}]{stickler2013theory}%
  \BibitemOpen
  \bibfield  {author} {\bibinfo {author} {\bibfnamefont {B.}~\bibnamefont
  {Stickler}},\ }\enquote {\bibinfo {title} {Theory and modeling of
  spin-transport on the microscopic and the mesoscopic scale},}\ \ (\bibinfo
  {publisher} {na},\ \bibinfo {year} {2013})\BibitemShut {NoStop}%
\bibitem [{\citenamefont {Datta}(2000)}]{datta2000nanoscale}%
  \BibitemOpen
  \bibfield  {author} {\bibinfo {author} {\bibfnamefont {S.}~\bibnamefont
  {Datta}},\ }\href@noop {} {\bibfield  {journal} {\bibinfo  {journal}
  {Superlattices and microstructures}\ }\textbf {\bibinfo {volume} {28}},\
  \bibinfo {pages} {253} (\bibinfo {year} {2000})}\BibitemShut {NoStop}%
\bibitem [{\citenamefont {Lee}\ \emph {et~al.}(1964)\citenamefont {Lee},
  \citenamefont {Logan}, \citenamefont {Batdorf}, \citenamefont {Kleimack},\
  and\ \citenamefont {Wiegmann}}]{lee1964ionization}%
  \BibitemOpen
  \bibfield  {author} {\bibinfo {author} {\bibfnamefont {C.}~\bibnamefont
  {Lee}}, \bibinfo {author} {\bibfnamefont {R.}~\bibnamefont {Logan}}, \bibinfo
  {author} {\bibfnamefont {R.}~\bibnamefont {Batdorf}}, \bibinfo {author}
  {\bibfnamefont {J.}~\bibnamefont {Kleimack}}, \ and\ \bibinfo {author}
  {\bibfnamefont {W.}~\bibnamefont {Wiegmann}},\ }\href@noop {} {\bibfield
  {journal} {\bibinfo  {journal} {Physical review}\ }\textbf {\bibinfo {volume}
  {134}},\ \bibinfo {pages} {A761} (\bibinfo {year} {1964})}\BibitemShut
  {NoStop}%
\bibitem [{\citenamefont {Conradi}(1972)}]{conradi1972distribution}%
  \BibitemOpen
  \bibfield  {author} {\bibinfo {author} {\bibfnamefont {J.}~\bibnamefont
  {Conradi}},\ }\href@noop {} {\bibfield  {journal} {\bibinfo  {journal} {IEEE
  Transactions on Electron Devices}\ }\textbf {\bibinfo {volume} {19}},\
  \bibinfo {pages} {713} (\bibinfo {year} {1972})}\BibitemShut {NoStop}%
\bibitem [{\citenamefont {Grant}(1973)}]{grant1973electron}%
  \BibitemOpen
  \bibfield  {author} {\bibinfo {author} {\bibfnamefont {W.}~\bibnamefont
  {Grant}},\ }\href@noop {} {\bibfield  {journal} {\bibinfo  {journal}
  {Solid-State Electronics}\ }\textbf {\bibinfo {volume} {16}},\ \bibinfo
  {pages} {1189} (\bibinfo {year} {1973})}\BibitemShut {NoStop}%
\bibitem [{\citenamefont {Kaneda}\ \emph {et~al.}(1976)\citenamefont {Kaneda},
  \citenamefont {Matsumoto},\ and\ \citenamefont {Yamaoka}}]{kaneda1976model}%
  \BibitemOpen
  \bibfield  {author} {\bibinfo {author} {\bibfnamefont {T.}~\bibnamefont
  {Kaneda}}, \bibinfo {author} {\bibfnamefont {H.}~\bibnamefont {Matsumoto}}, \
  and\ \bibinfo {author} {\bibfnamefont {T.}~\bibnamefont {Yamaoka}},\
  }\href@noop {} {\bibfield  {journal} {\bibinfo  {journal} {Journal of Applied
  Physics}\ }\textbf {\bibinfo {volume} {47}},\ \bibinfo {pages} {3135}
  (\bibinfo {year} {1976})}\BibitemShut {NoStop}%
\bibitem [{\citenamefont {Marshall}\ \emph {et~al.}(2008)\citenamefont
  {Marshall}, \citenamefont {Tan}, \citenamefont {Steer},\ and\ \citenamefont
  {David}}]{marshall2008electron}%
  \BibitemOpen
  \bibfield  {author} {\bibinfo {author} {\bibfnamefont {A.}~\bibnamefont
  {Marshall}}, \bibinfo {author} {\bibfnamefont {C.}~\bibnamefont {Tan}},
  \bibinfo {author} {\bibfnamefont {M.}~\bibnamefont {Steer}}, \ and\ \bibinfo
  {author} {\bibfnamefont {J.}~\bibnamefont {David}},\ }\href@noop {}
  {\bibfield  {journal} {\bibinfo  {journal} {Applied Physics Letters}\
  }\textbf {\bibinfo {volume} {93}},\ \bibinfo {pages} {111107} (\bibinfo
  {year} {2008})}\BibitemShut {NoStop}%
\bibitem [{\citenamefont {Marshall}\ \emph {et~al.}(2011)\citenamefont
  {Marshall}, \citenamefont {Ker}, \citenamefont {Krysa}, \citenamefont
  {David},\ and\ \citenamefont {Tan}}]{marshall2011high}%
  \BibitemOpen
  \bibfield  {author} {\bibinfo {author} {\bibfnamefont {A.~R.}\ \bibnamefont
  {Marshall}}, \bibinfo {author} {\bibfnamefont {P.~J.}\ \bibnamefont {Ker}},
  \bibinfo {author} {\bibfnamefont {A.}~\bibnamefont {Krysa}}, \bibinfo
  {author} {\bibfnamefont {J.~P.}\ \bibnamefont {David}}, \ and\ \bibinfo
  {author} {\bibfnamefont {C.~H.}\ \bibnamefont {Tan}},\ }\href@noop {}
  {\bibfield  {journal} {\bibinfo  {journal} {Optics express}\ }\textbf
  {\bibinfo {volume} {19}},\ \bibinfo {pages} {23341} (\bibinfo {year}
  {2011})}\BibitemShut {NoStop}%
\bibitem [{\citenamefont {Sun}\ \emph {et~al.}(2014)\citenamefont {Sun},
  \citenamefont {Maddox}, \citenamefont {Bank},\ and\ \citenamefont
  {Campbell}}]{sun2014record}%
  \BibitemOpen
  \bibfield  {author} {\bibinfo {author} {\bibfnamefont {W.}~\bibnamefont
  {Sun}}, \bibinfo {author} {\bibfnamefont {S.~J.}\ \bibnamefont {Maddox}},
  \bibinfo {author} {\bibfnamefont {S.~R.}\ \bibnamefont {Bank}}, \ and\
  \bibinfo {author} {\bibfnamefont {J.~C.}\ \bibnamefont {Campbell}},\ }in\
  \href@noop {} {\emph {\bibinfo {booktitle} {72nd Device Research
  Conference}}}\ (\bibinfo {organization} {IEEE},\ \bibinfo {year} {2014})\
  pp.\ \bibinfo {pages} {47--48}\BibitemShut {NoStop}%
\bibitem [{\citenamefont {Sun}\ \emph {et~al.}(2012)\citenamefont {Sun},
  \citenamefont {Lu}, \citenamefont {Zheng}, \citenamefont {Campbell},
  \citenamefont {Maddox}, \citenamefont {Nair},\ and\ \citenamefont
  {Bank}}]{sun2012high}%
  \BibitemOpen
  \bibfield  {author} {\bibinfo {author} {\bibfnamefont {W.}~\bibnamefont
  {Sun}}, \bibinfo {author} {\bibfnamefont {Z.}~\bibnamefont {Lu}}, \bibinfo
  {author} {\bibfnamefont {X.}~\bibnamefont {Zheng}}, \bibinfo {author}
  {\bibfnamefont {J.~C.}\ \bibnamefont {Campbell}}, \bibinfo {author}
  {\bibfnamefont {S.~J.}\ \bibnamefont {Maddox}}, \bibinfo {author}
  {\bibfnamefont {H.~P.}\ \bibnamefont {Nair}}, \ and\ \bibinfo {author}
  {\bibfnamefont {S.~R.}\ \bibnamefont {Bank}},\ }\href@noop {} {\bibfield
  {journal} {\bibinfo  {journal} {IEEE Journal of Quantum Electronics}\
  }\textbf {\bibinfo {volume} {49}},\ \bibinfo {pages} {154} (\bibinfo {year}
  {2012})}\BibitemShut {NoStop}%
\bibitem [{\citenamefont {Ker}\ \emph {et~al.}(2012)\citenamefont {Ker},
  \citenamefont {Marshall}, \citenamefont {Krysa}, \citenamefont {David},\ and\
  \citenamefont {Tan}}]{ker2012inas}%
  \BibitemOpen
  \bibfield  {author} {\bibinfo {author} {\bibfnamefont {P.~J.}\ \bibnamefont
  {Ker}}, \bibinfo {author} {\bibfnamefont {A.~R.}\ \bibnamefont {Marshall}},
  \bibinfo {author} {\bibfnamefont {A.~B.}\ \bibnamefont {Krysa}}, \bibinfo
  {author} {\bibfnamefont {J.~P.}\ \bibnamefont {David}}, \ and\ \bibinfo
  {author} {\bibfnamefont {C.~H.}\ \bibnamefont {Tan}},\ }in\ \href@noop {}
  {\emph {\bibinfo {booktitle} {2012 17th Opto-Electronics and Communications
  Conference}}}\ (\bibinfo {organization} {IEEE},\ \bibinfo {year} {2012})\
  pp.\ \bibinfo {pages} {220--221}\BibitemShut {NoStop}%
\bibitem [{\citenamefont {Beck}\ \emph {et~al.}(2001)\citenamefont {Beck},
  \citenamefont {Wan}, \citenamefont {Kinch},\ and\ \citenamefont
  {Robinson}}]{beck2001mwir}%
  \BibitemOpen
  \bibfield  {author} {\bibinfo {author} {\bibfnamefont {J.~D.}\ \bibnamefont
  {Beck}}, \bibinfo {author} {\bibfnamefont {C.-F.}\ \bibnamefont {Wan}},
  \bibinfo {author} {\bibfnamefont {M.~A.}\ \bibnamefont {Kinch}}, \ and\
  \bibinfo {author} {\bibfnamefont {J.~E.}\ \bibnamefont {Robinson}},\ }in\
  \href@noop {} {\emph {\bibinfo {booktitle} {Materials for Infrared
  Detectors}}},\ Vol.\ \bibinfo {volume} {4454}\ (\bibinfo {organization}
  {International Society for Optics and Photonics},\ \bibinfo {year} {2001})\
  pp.\ \bibinfo {pages} {188--197}\BibitemShut {NoStop}%
\bibitem [{\citenamefont {Beck}\ \emph {et~al.}(2004)\citenamefont {Beck},
  \citenamefont {Wan}, \citenamefont {Kinch}, \citenamefont {Robinson},
  \citenamefont {Mitra}, \citenamefont {Scritchfield}, \citenamefont {Ma},\
  and\ \citenamefont {Campbell}}]{beck2004hgcdte}%
  \BibitemOpen
  \bibfield  {author} {\bibinfo {author} {\bibfnamefont {J.~D.}\ \bibnamefont
  {Beck}}, \bibinfo {author} {\bibfnamefont {C.-F.}\ \bibnamefont {Wan}},
  \bibinfo {author} {\bibfnamefont {M.~A.}\ \bibnamefont {Kinch}}, \bibinfo
  {author} {\bibfnamefont {J.~E.}\ \bibnamefont {Robinson}}, \bibinfo {author}
  {\bibfnamefont {P.}~\bibnamefont {Mitra}}, \bibinfo {author} {\bibfnamefont
  {R.~E.}\ \bibnamefont {Scritchfield}}, \bibinfo {author} {\bibfnamefont
  {F.}~\bibnamefont {Ma}}, \ and\ \bibinfo {author} {\bibfnamefont {J.~C.}\
  \bibnamefont {Campbell}},\ }in\ \href@noop {} {\emph {\bibinfo {booktitle}
  {Infrared Detector Materials and Devices}}},\ Vol.\ \bibinfo {volume} {5564}\
  (\bibinfo {organization} {International Society for Optics and Photonics},\
  \bibinfo {year} {2004})\ pp.\ \bibinfo {pages} {44--53}\BibitemShut {NoStop}%
\bibitem [{\citenamefont {{Yu Ling Goh}}\ \emph {et~al.}(2005)\citenamefont
  {{Yu Ling Goh}}, \citenamefont {{Jo Shien Ng}}, \citenamefont {{Chee Hing
  Tan}}, \citenamefont {{Ng}},\ and\ \citenamefont {{David}}}]{InGaAs_random}%
  \BibitemOpen
  \bibfield  {author} {\bibinfo {author} {\bibnamefont {{Yu Ling Goh}}},
  \bibinfo {author} {\bibnamefont {{Jo Shien Ng}}}, \bibinfo {author}
  {\bibnamefont {{Chee Hing Tan}}}, \bibinfo {author} {\bibfnamefont {W.~K.}\
  \bibnamefont {{Ng}}}, \ and\ \bibinfo {author} {\bibfnamefont {J.~P.~R.}\
  \bibnamefont {{David}}},\ }\href {\doibase 10.1109/LPT.2005.857239}
  {\bibfield  {journal} {\bibinfo  {journal} {IEEE Photonics Technology
  Letters}\ }\textbf {\bibinfo {volume} {17}},\ \bibinfo {pages} {2412}
  (\bibinfo {year} {2005})}\BibitemShut {NoStop}%
\bibitem [{\citenamefont {Lenox}\ \emph {et~al.}(1998)\citenamefont {Lenox},
  \citenamefont {Yuan}, \citenamefont {Nie}, \citenamefont {Baklenov},
  \citenamefont {Hansing}, \citenamefont {Campbell}, \citenamefont
  {Holmes~Jr},\ and\ \citenamefont {Streetman}}]{InAlAs_random}%
  \BibitemOpen
  \bibfield  {author} {\bibinfo {author} {\bibfnamefont {C.}~\bibnamefont
  {Lenox}}, \bibinfo {author} {\bibfnamefont {P.}~\bibnamefont {Yuan}},
  \bibinfo {author} {\bibfnamefont {H.}~\bibnamefont {Nie}}, \bibinfo {author}
  {\bibfnamefont {O.}~\bibnamefont {Baklenov}}, \bibinfo {author}
  {\bibfnamefont {C.}~\bibnamefont {Hansing}}, \bibinfo {author} {\bibfnamefont
  {J.}~\bibnamefont {Campbell}}, \bibinfo {author} {\bibfnamefont
  {A.}~\bibnamefont {Holmes~Jr}}, \ and\ \bibinfo {author} {\bibfnamefont
  {B.}~\bibnamefont {Streetman}},\ }\href@noop {} {\bibfield  {journal}
  {\bibinfo  {journal} {Applied physics letters}\ }\textbf {\bibinfo {volume}
  {73}},\ \bibinfo {pages} {783} (\bibinfo {year} {1998})}\BibitemShut
  {NoStop}%
\bibitem [{\citenamefont {Ahmed}\ \emph
  {et~al.}(2018{\natexlab{b}})\citenamefont {Ahmed}, \citenamefont {Tan},
  \citenamefont {Zheng}, \citenamefont {Campbell},\ and\ \citenamefont
  {Ghosh}}]{ahmed2018apd}%
  \BibitemOpen
  \bibfield  {author} {\bibinfo {author} {\bibfnamefont {S.~Z.}\ \bibnamefont
  {Ahmed}}, \bibinfo {author} {\bibfnamefont {Y.}~\bibnamefont {Tan}}, \bibinfo
  {author} {\bibfnamefont {J.}~\bibnamefont {Zheng}}, \bibinfo {author}
  {\bibfnamefont {J.~C.}\ \bibnamefont {Campbell}}, \ and\ \bibinfo {author}
  {\bibfnamefont {A.~W.}\ \bibnamefont {Ghosh}},\ }in\ \href@noop {} {\emph
  {\bibinfo {booktitle} {2018 IEEE Photonics Conference (IPC)}}}\ (\bibinfo
  {organization} {IEEE},\ \bibinfo {year} {2018})\ pp.\ \bibinfo {pages}
  {1--2}\BibitemShut {NoStop}%
\bibitem [{\citenamefont {Piprek}(2013)}]{piprek2013semiconductor}%
  \BibitemOpen
  \bibfield  {author} {\bibinfo {author} {\bibfnamefont {J.}~\bibnamefont
  {Piprek}},\ }\enquote {\bibinfo {title} {Semiconductor optoelectronic
  devices: introduction to physics and simulation},}\ \ (\bibinfo  {publisher}
  {Elsevier},\ \bibinfo {year} {2013})\BibitemShut {NoStop}%
\bibitem [{\citenamefont {Shur}(1996)}]{shur1996handbook}%
  \BibitemOpen
  \bibfield  {author} {\bibinfo {author} {\bibfnamefont {M.~S.}\ \bibnamefont
  {Shur}},\ }\href@noop {} {\emph {\bibinfo {title} {Handbook series on
  semiconductor parameters}}},\ Vol.~\bibinfo {volume} {1}\ (\bibinfo
  {publisher} {World Scientific},\ \bibinfo {year} {1996})\BibitemShut
  {NoStop}%
\bibitem [{\citenamefont {Fawcett}\ \emph {et~al.}(1970)\citenamefont
  {Fawcett}, \citenamefont {Boardman},\ and\ \citenamefont
  {Swain}}]{FAWCETT19701963}%
  \BibitemOpen
  \bibfield  {author} {\bibinfo {author} {\bibfnamefont {W.}~\bibnamefont
  {Fawcett}}, \bibinfo {author} {\bibfnamefont {A.}~\bibnamefont {Boardman}}, \
  and\ \bibinfo {author} {\bibfnamefont {S.}~\bibnamefont {Swain}},\ }\href
  {\doibase https://doi.org/10.1016/0022-3697(70)90001-6} {\bibfield  {journal}
  {\bibinfo  {journal} {Journal of Physics and Chemistry of Solids}\ }\textbf
  {\bibinfo {volume} {31}},\ \bibinfo {pages} {1963} (\bibinfo {year}
  {1970})}\BibitemShut {NoStop}%
\bibitem [{\citenamefont {Yang}\ \emph {et~al.}(2020)\citenamefont {Yang},
  \citenamefont {Deng}, \citenamefont {Wei}, \citenamefont {Li},\ and\
  \citenamefont {Luo}}]{yang2020materials}%
  \BibitemOpen
  \bibfield  {author} {\bibinfo {author} {\bibfnamefont {Q.~L.}\ \bibnamefont
  {Yang}}, \bibinfo {author} {\bibfnamefont {H.~X.}\ \bibnamefont {Deng}},
  \bibinfo {author} {\bibfnamefont {S.~H.}\ \bibnamefont {Wei}}, \bibinfo
  {author} {\bibfnamefont {S.~S.}\ \bibnamefont {Li}}, \ and\ \bibinfo {author}
  {\bibfnamefont {J.~W.}\ \bibnamefont {Luo}},\ }\href@noop {} {\enquote
  {\bibinfo {title} {Materials design principles towards high hole mobility
  learning from an abnormally low hole mobility of silicon},}\ } (\bibinfo
  {year} {2020}),\ \Eprint {http://arxiv.org/abs/2011.02262} {arXiv:2011.02262
  [cond-mat.mtrl-sci]} \BibitemShut {NoStop}%
\bibitem [{\citenamefont {Palankovski}(2000)}]{Palankovski}%
  \BibitemOpen
  \bibfield  {author} {\bibinfo {author} {\bibfnamefont {V.}~\bibnamefont
  {Palankovski}},\ }\enquote {\bibinfo {title} {Simulation of heterojunction
  bipolar transistors},}\ \ (\bibinfo {year} {2000})\BibitemShut {NoStop}%
\bibitem [{\citenamefont {Ridley}(2013)}]{ridley2013quantum}%
  \BibitemOpen
  \bibfield  {author} {\bibinfo {author} {\bibfnamefont {B.~K.}\ \bibnamefont
  {Ridley}},\ }\href@noop {} {\emph {\bibinfo {title} {Quantum processes in
  semiconductors}}}\ (\bibinfo  {publisher} {Oxford university press},\
  \bibinfo {year} {2013})\BibitemShut {NoStop}%
\bibitem [{\citenamefont {Kodati}\ \emph {et~al.}(2021)\citenamefont {Kodati},
  \citenamefont {Lee}, \citenamefont {Guo}, \citenamefont {Jones},
  \citenamefont {Schwartz}, \citenamefont {Winslow}, \citenamefont {Pfiester},
  \citenamefont {Grein}, \citenamefont {Ronningen}, \citenamefont {Campbell},\
  and\ \citenamefont {Krishna}}]{AlInAsSb_expt2}%
  \BibitemOpen
  \bibfield  {author} {\bibinfo {author} {\bibfnamefont {S.~H.}\ \bibnamefont
  {Kodati}}, \bibinfo {author} {\bibfnamefont {S.}~\bibnamefont {Lee}},
  \bibinfo {author} {\bibfnamefont {B.}~\bibnamefont {Guo}}, \bibinfo {author}
  {\bibfnamefont {A.~H.}\ \bibnamefont {Jones}}, \bibinfo {author}
  {\bibfnamefont {M.}~\bibnamefont {Schwartz}}, \bibinfo {author}
  {\bibfnamefont {M.}~\bibnamefont {Winslow}}, \bibinfo {author} {\bibfnamefont
  {N.~A.}\ \bibnamefont {Pfiester}}, \bibinfo {author} {\bibfnamefont {C.~H.}\
  \bibnamefont {Grein}}, \bibinfo {author} {\bibfnamefont {T.~J.}\ \bibnamefont
  {Ronningen}}, \bibinfo {author} {\bibfnamefont {J.~C.}\ \bibnamefont
  {Campbell}}, \ and\ \bibinfo {author} {\bibfnamefont {S.}~\bibnamefont
  {Krishna}},\ }\href {\doibase 10.1063/5.0039399} {\bibfield  {journal}
  {\bibinfo  {journal} {Applied Physics Letters}\ }\textbf {\bibinfo {volume}
  {118}},\ \bibinfo {pages} {091101} (\bibinfo {year} {2021})},\ \Eprint
  {http://arxiv.org/abs/https://doi.org/10.1063/5.0039399}
  {https://doi.org/10.1063/5.0039399} \BibitemShut {NoStop}%
\bibitem [{\citenamefont {Xie}\ \emph {et~al.}(2012)\citenamefont {Xie},
  \citenamefont {Xie}, \citenamefont {Tozer},\ and\ \citenamefont
  {Tan}}]{AlAsSb_expt2}%
  \BibitemOpen
  \bibfield  {author} {\bibinfo {author} {\bibfnamefont {J.}~\bibnamefont
  {Xie}}, \bibinfo {author} {\bibfnamefont {S.}~\bibnamefont {Xie}}, \bibinfo
  {author} {\bibfnamefont {R.~C.}\ \bibnamefont {Tozer}}, \ and\ \bibinfo
  {author} {\bibfnamefont {C.~H.}\ \bibnamefont {Tan}},\ }\href {\doibase
  10.1109/TED.2012.2187211} {\bibfield  {journal} {\bibinfo  {journal} {IEEE
  Transactions on Electron Devices}\ }\textbf {\bibinfo {volume} {59}},\
  \bibinfo {pages} {1475} (\bibinfo {year} {2012})}\BibitemShut {NoStop}%
\end{thebibliography}
\end{document}